\documentclass[aps,pra,reprint,floatfix]{revtex4-1}
\usepackage{amsmath}
\usepackage{amssymb}
\usepackage{graphicx}
\usepackage{bm}

\newcommand{\orderof}{\ensuremath{\mathcal{O}}}
\newcommand{\abs}[1]{\left| #1 \right|}
\newcommand{\bvec}[1]{\ensuremath\mathbf{ #1 }}

\begin{document}

\title{Conical diffraction and the dispersion surface of hyperbolic 
metamaterials}
\author{K. E. Ballantine}
\author{J. F. Donegan}
\author{P. R. Eastham}
\affiliation{School of Physics and CRANN, Trinity College Dublin, Dublin 2, 
Ireland}

\begin{abstract}
  Hyperbolic metamaterials are materials in which at least one
  principal dielectric constant is negative. We describe the
  refractive index surface, and the resulting refraction effects, for
  a biaxial hyperbolic metamaterial, with principal dielectric
  constants $\epsilon_1<0$, $0<\epsilon_2\neq\epsilon_3$.  In this
  general case the two sheets of the index surface intersect
  forming conical singularities. We derive the ray description of
  conical refraction in these materials, and show that it is
  topologically and quantitatively distinct from conical refraction in
  a conventional biaxial material. We also develop a wave optics
  description, which allows us to obtain the diffraction patterns
  formed from arbitrary beams incident close to the optic axis. The
  resulting patterns lack circular symmetry, and hence are
  qualitatively different from those obtained in conventional,
  positive index materials.
\end{abstract}
\maketitle

\section{Introduction}
Hyperbolic metamaterials (HMMs), materials which have a negative dielectric 
constant in at least one direction, are attracting attention due to 
their interesting physics and myriad applications. They can be 
manufactured relatively simply from alternating layers of metal and 
dielectric, or by embedding metal rods in a dielectric background~\cite{
Cortes12,Podolskiy05}. HMMs have recently been shown to have unique 
properties, described by effective medium theory~\cite{Kidwai12}, including 
a broadband infinite density of states~\cite{Jacob12}, arbitrarily large 
values of the wavevector~\cite{Yang12}, and negative refraction~\cite{
Podolskiy05,Smith04}. This has led to many proposed applications, from 
imaging~\cite{Jacob06,Liu07}, sensing~\cite{Kabashin09}, and wave 
guiding~\cite{Govyadinov06,He12} to information processing~\cite{Wurtz11}. 

The most common HMMs considered are uniaxial materials for which
$\epsilon_ 1< 0<\epsilon_2=\epsilon_3$ where $\epsilon_i$ are the
principal dielectric constants. This leads to a hyperboloid
isofrequency surface (refractive index surface) for the extraordinary
ray. The change in topology from an ellipsoid to a hyperboloid is
responsible for many of the important properties of these 
materials~\cite{Krishna12}. The general case, however, is a biaxial HMM, where
$\epsilon_1<0< \epsilon_2<\epsilon_3$. Such a material could be
realized as layers of metal and dielectric, where the dielectric
material has uniaxial isotropy in the plane, or as rods of metal
embedded in a dielectric with different rod spacings in the $x$ and
$y$ directions~\cite{Elser06}. The isofrequency surface for the
extraordinary ray is then an asymmetric hyperboloid~\cite{Sun13}.

In this paper we present the full two-sheeted isofrequency surface of a 
HMM, which describes the propagation of both the ordinary and the 
extraordinary rays with orthogonal polarizations, 
and show that it contains conical singularities. These singularities are 
degenerate points where the two sheets intersect at a point in $k$-space. 
Similar conical singularities occur in 
conventional biaxial materials, i.e., $0< \epsilon_1<\epsilon_2< \epsilon_3$
~\cite{Berry03,Berry04}, and lead to the phenomenon of conical refraction, 
in which a beam of light is 
refracted into two concentric hollow cones~\cite{Berry04,Portigal69}. We 
describe these intersections in the case of a HMM, and derive a geometrical 
optics description of refraction for 
rays with wavevector close to the degeneracy, including establishing the 
polarization and the Poynting vector, or energy flow. This predicts 
refraction into two intersecting rather than 
concentric cones, an effect topologically distinct from that in a 
conventional biaxial crystal and completely lacking from a uniaxial HMM. We 
then extend this theory to develop a paraxial 
wave optics description of the propagation of light through these 
materials. This allows us to calculate the diffraction patterns formed from 
arbitrary beams incident on a biaxial HMM 
close to the optic axis. We find these patterns to be qualitatively 
different from those obtained in positive index materials, in particular 
lacking circular symmetry.

These conical singularities are, in some respects, similar to the Dirac 
points~\cite{Wallace47} that are of growing importance in solid-state 
physics. These points, where bands cross linearly at a particular frequency 
and wavevector, are best known in graphene~\cite{Novoselov04,Castro09}. 
Graphene has attracted huge theoretical and applied interest~\cite{
Geim07,Cooper11,Sarma11}, with many new features attributable to the linear 
dispersion near a Dirac point, which means that the low-energy excitations 
are massless chiral Dirac fermions~\cite{Castro09}. They thus provide a 
model of quantum electrodynamics with the limiting speed given by the Fermi 
velocity rather than the speed of light~\cite{Novoselov05,Castro09}. They 
also lead to effects such as the anomalous integer quantum Hall effect~\cite
{Gusynin05,Zhang05}, and mean that electrons are immune to localization, 
propagating over large distances without scattering~\cite{Lee85,Novoselov04}
.  Tilted Dirac cones, which are not circularly symmetric around the 
degenerate wavevector, are similar to the skewed-cone intersections 
reported here, and have previously been predicted in mechanically deformed 
graphene~\cite{Goerbig08}. Dirac points in optical systems have been found 
in photonic crystals, as a result of the same lattice symmetry~\cite{
Plihal91,Peleg07}, or in materials with a frequency dependent permittivity, 
which may pass through zero at a particular frequency leading to a 
degeneracy~\cite{Wang09,Huang11,LSun13}.

In these cases, however, a degeneracy occurs at a particular
frequency, due to fine tuning the frequency to match the sublattice
periodicity, or to match a zero of the frequency dependent dielectric
constant. At other nearby frequencies there is generally no
singularity. In contrast, biaxial materials have conical singularities
in the isofrequency surface in $k$-space, which is directly comparable
to a Fermi surface.  The presence of these singularities depends on
the symmetry of orthogonal polarizations in a crystal, and does not
rely on fine tuning of any parameter. In particular, we show that they
occur within effective medium theory, and argue that their presence is
required on topological grounds. Since this implies that they occur
over a finite range of frequencies they correspond to line, rather
than point, degeneracies in the dispersion relation (which describes a
three dimensional surface in the four dimensional space of
$\omega$ and $\bvec{k}$).

The remainder of this article is structured as follows. In 
Sec.~\ref{sec:disp} we describe the two-sheeted dispersion surface 
in a biaxial HMM 
and compare it to the case of positive $\epsilon$.  In Sec.~\ref{sec:ray} 
we derive the ray optics description of refraction, for incident rays with 
initial wavevector close to the optic axis, in a biaxial HMM. In particular, 
we present the polarization and Poynting vector, i.e. the direction of 
energy flow, of the refracted rays.  In Sec.~\ref{sec:loss} we extend the 
theory to include small absorption in the material, and show explicitly 
that the conical intersections persist. In Sec.~\ref{sec:diff} we develop a 
wave-optics description of propagation near the optic axis of a biaxial 
HMM, and present the diffraction pattern formed with a Gaussian input 
beam.  In Sec.~\ref{sec:disc} we discuss further the connection between 
conical singularities in optics and singularities in solid-state 
bandstructures. We make an explicit connection between the conical 
singularities described here and Dirac points by reformulating the 
diffraction theory in terms of the paraxial wave equation. Finally, in Sec.~
\ref{sec:conc} we summarize our conclusions.

\section{Dispersion surfaces}
\label{sec:disp}

We can describe a nano-structured metamaterial in the effective medium 
theory by a three-dimensional dielectric tensor $\epsilon_{ij}$ or by the 
principal dielectric constants, $\epsilon_i$, which are its components in 
the frame in which it is diagonal~\cite{Born99}. Effective medium theory 
describes the sub-wavelength patterning of different materials by an 
average, anisotropic dielectric tensor according to the Maxwell-Garnett 
formulas~\cite{Niklasson81}.  Plane wave solutions to Maxwell's equations 
in the medium lead to the Fresnel equation for the refractive index, 
\begin{equation}\label{Fresnel}   
\sum_i\frac{\epsilon_i\eta_i^2}{n^2- \epsilon_i}=0, \end{equation} 
where $\bvec{\eta}$ is a unit vector in the direction of 
the wavevector $\bvec{k}$~\cite{Born99}. The two solutions for $n ^2$ for a 
given direction $\bvec{\eta}$ form a two-sheeted dispersion surface~\cite{
Born99}, also known as an isofrequency surface or refractive index surface. 
At a fixed frequency, these surfaces give the phase velocity, or 
equivalently the wavevector magnitude, in the medium, for a given 
wavevector direction. The ray or energy flow direction will be orthogonal 
to the dispersion surface at the point defined by that wavevector~\cite{
Landau84}. In the following we assume without loss of generality that $
\epsilon_1<\epsilon_2<\epsilon_3$.

\begin{figure}[htbp]
 \centering
 \includegraphics[width=\columnwidth]{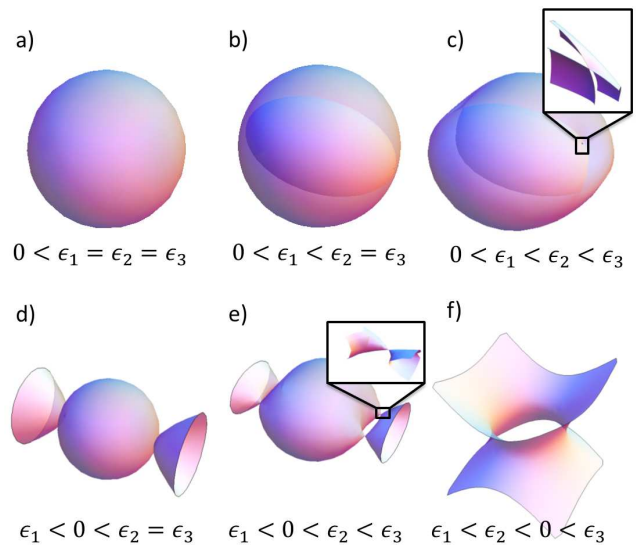}\\
  \caption{(color online) Isofrequency surfaces for various effective index 
materials: a) isotropic b) uniaxial c) biaxial d) uniaxial-hyperbolic e) 
biaxial-hyperbolic type 1 f) biaxial-hyperbolic type 2. Shading is for 
perspective only. Additional cases not shown include 
$\epsilon_1=\epsilon_2<0<\epsilon_3$, which is identical to f) but with 
circular cross-sections, and $\epsilon_3<0$, in which case there are no real 
solutions. These surfaces are polar plots of refractive index as a function of 
ray direction $\bvec{\eta}$. In the case of (b) and (d) the surfaces intersect 
at two points, at which they are parallel. In the case of (c) and (e) the 
surfaces have four conical intersections. Insets in (c) and (e) show cutaway 
close-ups of the intersection points. Cuts through these intersections 
are presented in Fig.~\ref{fig:trans}.\label{fig:surfaces}}
\end{figure}

Figure~\ref{fig:surfaces} shows sections of the dispersion surfaces for a 
variety of materials. These surfaces are polar plots where the radial 
distance represents the refractive index experienced by a ray propagating 
in that direction in $k$-space. Equivalently, they are three-dimensional 
cuts of the full four dimensional space of $\omega$ and $\bvec{k}$, taken 
at a constant $\omega$. In the approximation where the dielectric constants 
depend weakly on frequency, these surfaces will simply contract or expand 
as $\omega$ is decreased or increased respectively, meaning the critical 
points will trace out lines. Outside of this approximation the dispersion 
surface will change shape but the basic features will remain until the 
dielectric constants cross each other or zero. Hence assuming a smooth 
dependence on frequency there will always be a continuous range of 
frequencies for which these singularities exist.

The classical cases, $0<\epsilon_i$, are shown in the first row, and are 
the subject of conventional crystal optics. The surfaces have positive 
curvature and finite area. The hyperbolic cases, $\epsilon_1<0$ shown in 
the second row, are the result of nano-structured materials which have 
properties not found in nature at optical frequencies. They have dispersion 
surfaces which are unbounded in $\abs{k}$ at any frequency, and feature 
both positive and negative curvature~\cite{Cortes12}.   

The possible classical materials fall into three categories. 
Figure~\ref{fig:surfaces}(a) shows an isotropic material 
which has a single, spherical dispersion surface. 
Once isotropy is broken, the surface splits into two as 
the two orthogonal polarizations experience different dielectric constants. 
For a uniaxial material, with two indices equal, these surfaces intersect 
at two points, along a single optic axis as shown in Fig.~\ref{fig:surfaces}
(b). However the surfaces are parallel at the degenerate points, and so the 
normals remain well defined~\cite{Hecht02}. For a biaxial crystal, shown in 
Fig.~\ref{fig:surfaces}(c), rotational symmetry is broken completely. The 
surfaces intersect at four points along two optic axes. The gradient of the 
surfaces is singular at the degenerate points and the normal is not well 
defined. 

These singularities lead to the unique phenomenon of conical 
refraction~\cite{Berry03}. For a general angle of incidence in an anisotropic
medium, the two orthogonal polarizations of an incident ray are 
refracted into two rays with different wavevectors, called the ordinary
and extraordinary rays. In conical refraction, when the incident
wavevector coincides with the optic axis, 
the two orthogonally polarized incident rays are refracted into two 
concentric cones which contain all polarizations at different points around 
each cone~\cite{Portigal69,Born99}.

\begin{figure}[htbp]
 \centering
 \includegraphics[width=\columnwidth]{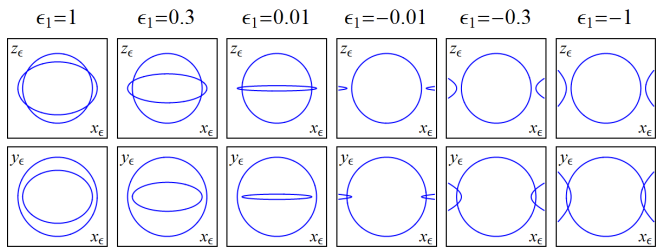}\\
  \caption{(color online) The transition from biaxial to biaxial-hyperbolic 
type 1 material as $\epsilon_1$ passes through 0. One of the dispersion 
surfaces changes topology from an ellipsoid to a hyperboloid. The 
intersection points move from the $x_\epsilon$-$z_\epsilon$ plane to the 
$x_\epsilon$-$y_\epsilon$  plane. The first row shows the surfaces in the 
$x_\epsilon$-$z_\epsilon$ plane ($y_\epsilon$=0). The second row shows 
the surfaces in the
$x_\epsilon$-$y_\epsilon$ plane ($z_\epsilon$=0).
  \label{fig:trans}}
\end{figure}

When one of the dielectric constants becomes negative, leading to a
hyperbolic metamaterial, there is a topological transition of one of
the surfaces, from an ellipsoid to a hyperboloid. 
Figure~\ref{fig:surfaces}(d) shows a uniaxial HMM. The surfaces again
intersect at two points where they are parallel. In the case of a
biaxial HMM, shown in Fig.~\ref{fig:surfaces}(e), linear crossings
occur. The hyperboloid and the ellipsoid intersect at four degenerate
points. We describe for the first time these conical singularities in
the dispersion surface of a biaxial HMM, and their associated
refraction and diffraction effects. In the final case, where two of the
three indices are negative, Fig.~\ref{fig:surfaces}(f), there is again
a single dispersion surface which is a type two hyperboloid~\cite{Cortes12} 
with no singularities. This single dispersion surface
describes one polarization which can propagate in the material. For
the orthogonal polarization the material is metallic, and absorbing,
hence there is no second real solution to the Fresnel equation.

In both Fig~\ref{fig:surfaces}(b) and (d) the two sheets have a
quadratic degeneracy. Including the perturbation
$\epsilon_2\neq\epsilon_3$ will clearly either open a gap or cause the
quadratic intersection to split into two linear intersections, in line
with general band theory. If a gap were to open, however, it would
leave at least one closed surface which described the propagation of a
different linear polarization at each point. The field of polarization
directions described by this surface would form a tangential vector
field on a closed two-dimensional surface. This is forbidden by the
hairy ball theorem, unless the linear polarization vanishes at least
once. Comparing with the Poincar\'{e} sphere representation for the
polarization, we see that such points, if they occurred, would
correspond to points with circular polarization. However, in the
presence of chiral symmetry the two circular polarizations cannot have
different refractive indices, so that there cannot be a gap at these
points. Thus, in the presence of chiral symmetry, the existence of
conical singularities in the isofrequency surface is required on
topological grounds. In its absence, however, a gap does indeed
appear~\cite{Gao14}.

The transition from a conventional biaxial material to a biaxial type 1 HMM is
shown in Fig.~\ref{fig:trans} as $\epsilon_1$ goes from positive to negative.
As rotational symmetry in the $y_\epsilon$-$z_\epsilon$ plane is 
broken (note we use the subscript $\epsilon$ to denote the basis in 
which $\epsilon$ is diagonal), the 
degenerate points are free to move around the 
$x_\epsilon$ axis as $\epsilon_1$
varies. The points start in the $x_\epsilon$-$z_\epsilon$ plane and move 
closer to the $x_\epsilon$ axis as $\epsilon_1\rightarrow0$. Then as the 
topological transition
occurs the critical points change direction and move away 
from the $x_\epsilon$
axis into the $x_\epsilon$-$y_\epsilon$ plane.

The topological transition between the conical singularities of
positive and negative index materials can be seen by calculating the
solutions to the Fresnel equation~\eqref{Fresnel} which are
degenerate. We find two sets of solutions
\begin{equation}\label{degenpoints1} \begin{aligned} 
\eta_1 &=\pm\sqrt{\frac{\epsilon_3 \left(\epsilon_2-\epsilon_1\right)}
{\epsilon_2 \left(\epsilon_3-\epsilon_1\right)}} \\ \eta_2 &=0 \\ \eta_3 
&=\pm\sqrt{\frac{ \epsilon_1 \left(\epsilon_3
-\epsilon_2\right)}{\epsilon_2\left(\epsilon_3- \epsilon_1\right)}}  
\end{aligned}\end{equation} 
and 
\begin{equation}\label{degenpoints2} \begin{aligned} \eta_1 &=\pm\sqrt{\frac
{\epsilon_2 \left( \epsilon_3-\epsilon_1\right)}{\epsilon_3\left(\epsilon_2-
\epsilon_1\right)}} \\ \eta_2 &=\pm\sqrt{\frac{-\epsilon_1 \left(\epsilon_3-
\epsilon_2\right)}{ \epsilon_3\left(\epsilon_2-\epsilon_1\right)}} \\ \eta_
3 &=0. \end{aligned} \end{equation} The first solution 
Eq. \eqref{degenpoints1} is real, and therefore physical, 
when all the $\epsilon_i$ 
are positive. As $\epsilon_1$ becomes negative $\eta_3$ in 
Eq. \eqref{degenpoints1} becomes imaginary. The second 
solution Eq.~\eqref{degenpoints2} 
then becomes the real, physically relevant, $ \bvec{\eta}$. In this way 
the transition through $\epsilon_1=0$ separates topologically distinct sets 
of degenerate solutions.

Figure~\ref{fig:sections} shows the cross-sections of the dispersion 
surfaces at the degenerate points, in the case of a conventional biaxial 
crystal and a biaxial hyperbolic material. For a conventional material, 
both surfaces have similar curvature. The normals to the surfaces close to 
the optic axis, i.e., the axis which passes through one of the degenerate 
points, are shown. These normals indicate the direction of refraction for 
rays which approximately coincide with the optic axis. In the positive $
\epsilon$ case, one points close to the optic axis while the other points 
away from the $x_\epsilon$ axis. In the case of a biaxial HMM the surfaces 
have opposite curvature. This leads to one of the normals pointing towards 
the $x_\epsilon$ axis. When the full two dimensional surface is considered, 
the normals shown here contribute to a cone which is skewed away from the 
optic axis, in a different direction in each case. In 
Fig.~\ref{fig:sections}(b), one of the normals 
points downwards, below the horizontal. If the 
material is cut so the interface is the $y_\epsilon$-$z_\epsilon$ plane, 
i.e. the normal is parallel to the $x_\epsilon$ axis, then this results in 
part of the cone being refracted back on the same side of the normal to the 
incoming ray, a phenomenon sometimes known as negative refraction. However 
this term is also used to refer to negative phase velocity, which is not 
present in this case.
 
\section{Geometrical Optics}
\label{sec:ray}

\begin{figure}[htbp]
 \centering
 \includegraphics[width=\columnwidth]{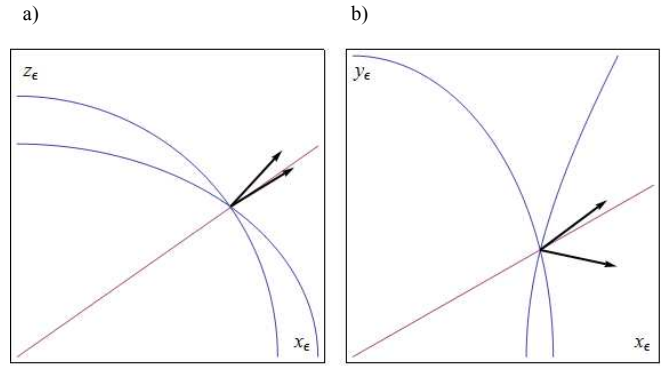}\\
  \caption{(color online) Cross-sections of the isofrequency surfaces 
through the degenerate points for  (a)  a conventional biaxial material 
and (b) a hyperbolic biaxial material.   The optic axis is shown by the 
straight line and the approximate normals to the surfaces for a 
$\bvec{k}$ vector passing close to this axis are shown by the arrows, and are
suggestive of the   expected conical refraction. In the hyperbolic case 
the cone points towards rather than away from the $x_\epsilon$- axis.
\label{fig:sections}}
\end{figure}

\begin{figure}[htbp]
 \centering
 \includegraphics[width=\columnwidth]{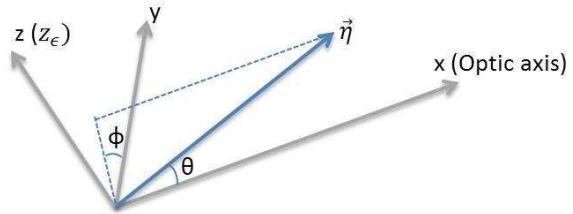}\\
  \caption{(color online) The coordinate system used to describe 
refraction near the optic axis in a biaxial HMM. The $x$ axis 
corresponds to the optic axis   through the direction given by Eq. 
\eqref{degenpoints2} while the $z$ axis corresponds to the 
$z_\epsilon$ axis. $\theta$ is the angular displacement of the ray from the
optic axis while $\phi$ is the azimuthal angle of the ray in the 
transverse plane.
  \label{fig:coords}}
\end{figure}

We now turn to describing the refraction of light incident on a
biaxial HMM, when the incident wavevector lies close to the optic
axis, as shown in Fig.~\ref{fig:sections}. To achieve this we
calculate the refractive index surface experienced by the ray and the
resulting Poynting vector of the refracted ray.  We describe the ray
by polar coordinates in a frame where the $x$ axis coincides with the
optic axis, and the $z$ axis coincides with the $z_\epsilon$ axis, as
illustrated in Fig.~\ref{fig:coords}.  $\theta$ is the angle between
the ray and the optic axis, while $\phi$ is the azimuthal angle from
the $y$ axis in the $y$-$z$ (transverse) plane. Expressing $\bvec{\eta}$
in terms of $\theta$ and $\phi$ and solving Eq. \eqref{Fresnel} we find
the refractive index to first order in $\theta$
is \begin{equation}\label{n2}
  n^2=\epsilon_3-\theta\epsilon_\delta\left(\cos{
      \phi}\pm1\right) \end{equation} where \begin{equation}\label{ed}
  \epsilon_
  \delta=\epsilon_3\sqrt{\frac{\left(\epsilon_3-\epsilon_1\right)
      \left(\epsilon _2-\epsilon_3\right)}{\epsilon_1
      \epsilon_2}} \end{equation} is a measure of the anisotropy of
the medium. The surface described by Eq. \eqref{n2} consists of two cones
touching at their points, which is the linear approximation to the
surface portrayed in Fig.~\ref{fig:surfaces}(e) around one of the
intersection points.  Furthermore, we find the polarization of the two
refracted rays is \begin{equation}\label{polarization}
  \frac{D_z}{D_y}= \frac{\sin{\phi}}{\cos{\phi}\pm1} \end{equation}
where $\bvec{D}$ is the electric displacement field.

The results Eqs. \eqref{n2} and \eqref{polarization} describe the 
refractive index experienced by an incoming ray. A ray which comes from an 
azimuthal angle $\phi$ can be decomposed into the two orthogonal 
polarizations given by Eq. \eqref{polarization}. These two polarizations 
experience the refractive indices given by Eq. \eqref{n2}. The 
polarizations are independent of $\theta$, as long as $ \theta$ is small. 
Thus for any ray not exactly coincident with the optic axis, there are two 
distinct polarization modes. As $\phi$ varies, the direction of 
polarization described by a given dispersion surface rotates, so that a ray 
with one linear polarization and azimuthal angle $\phi$ undergoes the same 
refraction as a ray with the orthogonal polarization and azimuthal angle $
\phi+180^\circ$. However Eq. \eqref{polarization} is undefined when $\theta 
=0$.  Hence there is also a polarization degeneracy at the conical 
singularity where all polarizations experience the same refractive index.

Equation \eqref{n2} differs from the usual case of conical refraction
in a biaxial crystal in two noteworthy ways. Firstly $\epsilon_3$ plays
the role of the average dielectric constant, despite being the largest
of the three indices, while for a conventional biaxial crystal the
median index $\epsilon_2$ plays this role. Secondly, the parameter
$\epsilon_\delta$ depends on $\sqrt{\epsilon_3- \epsilon_1}$, which is
a large parameter since $\epsilon_1$ is negative. In the conventional,
$\epsilon_i>0$, case of conical refraction the corresponding form is
$\epsilon_{\delta}= \epsilon_2\sqrt{\left(\epsilon_2-\epsilon_1\right)
  \left(\epsilon_3-\epsilon_2\right)/ \epsilon_1\epsilon_3}$, which is
usually small. The polarization modes given by Eq. \eqref{polarization} are
identical to the positive $\epsilon$ case. Thus we do not expect the
polarization profiles generated by conical refraction and diffraction
to change.

\begin{figure}[htbp]
 \includegraphics[width=\columnwidth]{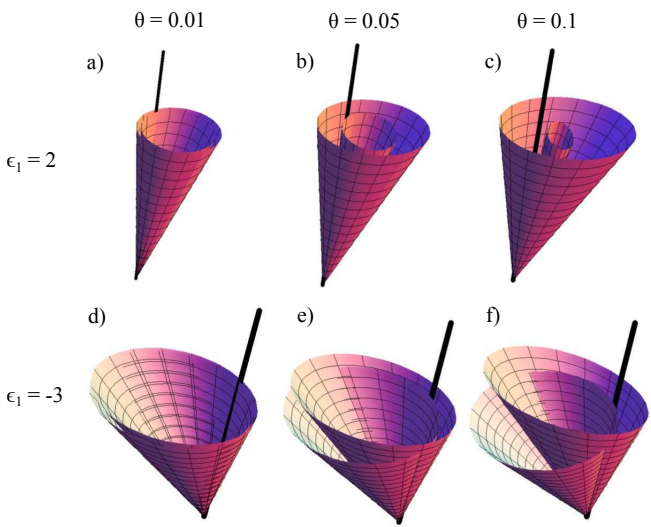}\\
  \caption{(color online) The loci of the Poynting vector of the 
two modes in a conventional biaxial material   and a biaxial hyperbolic 
metamaterial, for wavevectors making angles $\theta$ and $\phi$ to the 
optic axis, as $\phi$ varies from 0 to $2\pi$.   In the conventional case 
the cones are concentric, while in the hyperbolic case they intersect. For 
$\theta\rightarrow0$ the cones   are degenerate. As $\theta$ increases they 
move further apart.   Parameters used are $\epsilon_2=3$, $\epsilon_3=4$, 
top row; $\epsilon_1=2$ a) $\theta=0.01$, b) $\theta=0.05$, c) $\theta=0.1$
   and second row; $\epsilon_1=-3$ d) $\theta=0.01$, e) $\theta=0.05$, f) 
   $\theta=0.1$.   The solid black line indicates the optic axis, while the 
shading is   for perspective only.  
\label{fig:cones}}
\end{figure}

We now calculate the Poynting vector using Eqs. \eqref{n2} and 
\eqref{polarization} for the two orthogonal polarizations associated with each
incident wavevector. The Poynting vector is, up to an overall constant, given 
by
\begin{equation} \bvec{P}=\bvec{E}^*\times\bvec{H}.\end{equation}
$\bvec{E}$ and $\bvec{H}$ can be expressed in terms of $D_z$ and $D_y$, 
given by Eq.\eqref{polarization}, using Maxwell's equations and 
the constitutive relations. The result, 
\begin{equation}\label{Poynting}\begin{aligned} 
P_x &= \frac {1}{\epsilon_3^{3/2}}+\theta \frac{\epsilon_\delta}
{\epsilon_3^{5/2}}\left(\cos{\phi}\pm1\right) \\ P_y &= \frac{\epsilon_\delta}
{2\epsilon_3^{5/2}}\left(1\pm\cos{\phi}\right)+\frac{1}{\sqrt{ \epsilon_3}}
\theta\left[\pm\frac{\epsilon^2_\delta}{4\epsilon_3^3}
\left(\cos{ \phi}\pm1\right)^2\right.\\ &\left. +\frac{1}{2}\left(\frac{1}{
\epsilon_1}+ \frac{1}{\epsilon_2}\right )\left(\cos{\phi}\pm1\right)\mp\frac{1
}{\epsilon_3} \right] \\ P_z &= \pm\frac{\epsilon_\delta}{2\epsilon_3^{5/2}} 
\sin{\phi}+ \frac{1}{\sqrt{\epsilon_3}}\theta\left[\frac{\epsilon^2_ \delta}{
4\epsilon_3^3 }\left(\cos{\phi}\pm1\right)\sin{\phi}\right.\\ &\left.+\frac{1
}{2}\left(\frac {1}{\epsilon_1}+\frac{1}{\epsilon_2}\right) \sin{\phi}\right],
\end{aligned} \end{equation} is compared with the $\epsilon_i>0$ case in 
Fig.~\ref{fig:cones} for three values of $\theta$.

Equations \eqref{polarization} and  \eqref{Poynting} together describe the 
refraction of an incoming ray with wavevector at a small angle $\theta$ to the 
optic axis, 
and an azimuthal angle $\phi$ in the perpendicular plane. As $\phi$ is 
varied, the resulting rays sweep out two intersecting cones while the 
polarization component which is refracted into each cone also varies. For $
\theta=0$ a single ray of any polarization is refracted into a complete cone, 
containing all polarizations. However any realistic incoming beam will be a 
superposition of rays with the $\theta=0$ ray contributing an infinitesimal 
amount to the resulting pattern~\cite{Portigal69}.

Figure~\ref{fig:cones} shows the loci of the Poynting vectors at different 
fixed angles $\theta$ as the azimuthal angle $\phi$ is varied, for a biaxial 
conventional material and a biaxial HMM. This is indicative of the paths taken
by refracted rays in the material. 
The figures show that the usual result of two concentric 
cones~\cite{Portigal69} changes to the 
topologically distinct case of two intersecting cones. At $\theta\approx0$ 
the cones are degenerate, and skewed away from the optic axis. The degeneracy 
is clear from Eq. \eqref{Poynting}. For $\theta=0$ the terms which depend 
on $\phi$ take the same value for one mode at a given $\phi$ as for the other 
mode at $\phi+\pi$.  As $\theta$ increases, the cones move in opposite 
directions along the $y$ axis, so that they intersect and for large enough $
\theta$ will separate entirely. We note that this is due to a particular term 
in the Poynting vector, Eq. \eqref{Poynting}, \begin{equation}\label{py} 
P_y\propto 
\ldots +\theta\left[\frac{1}{2}\left(\frac{1}{\epsilon_1}+\frac{1}{\epsilon_2
}\right)\left(\cos{\phi}\pm1\right)\mp\frac{1}{\epsilon_3}\right] 
\end{equation} which is the dominant term for the movement of the cones as
$\theta$ increases. For $\epsilon_1\approx-\epsilon_2$, the first 
term in Eq. \eqref{py} is 
small, and so the two modes have terms $\approx \mp 1/\epsilon_3$ in $P_y$ of
opposite sign with little 
dependence on $\phi$. This means the entire cones will move in opposite 
directions as $\theta$ increases. There is a corresponding term in the 
conventional case, but there if $\epsilon_1\approx\epsilon_2\approx\epsilon_3$
it is the constant terms $\pm 1/2\epsilon_1 \pm1/2 \epsilon_2\mp1/\epsilon_3$
which approximately cancel, leaving a term which is 
dominated by $\cos{\phi}$. Thus the center of the cones do not move in this 
case. 

\section{Absorption}
\label{sec:loss}

So far it has been assumed that although the permittivity may be negative it 
will always be real. Since hyperbolic metamaterials contain a large proportion
of metal, they will always have some absorption, leading to an imaginary part
of the effective permittivity. Although metals generally have high absorption,
it is possible to design hyperbolic metamaterials with a small imaginary part
of $\epsilon$ over a range of frequencies~\cite{Kidwai12}. Nevertheless it is
important to 
consider how losses will affect the basic theory. 
Previous figures have plotted the real solutions of the Fresnel equation. In 
directions in which only one real solution exists, the other solution is 
completely imaginary and thus evanescent. When the permittivitty is complex, 
all solutions are complex, and represent waves which travel with some 
absorption, which depends on the size of the imaginary component.

\begin{figure}[htbp]
 \centering
 \includegraphics[width=\columnwidth]{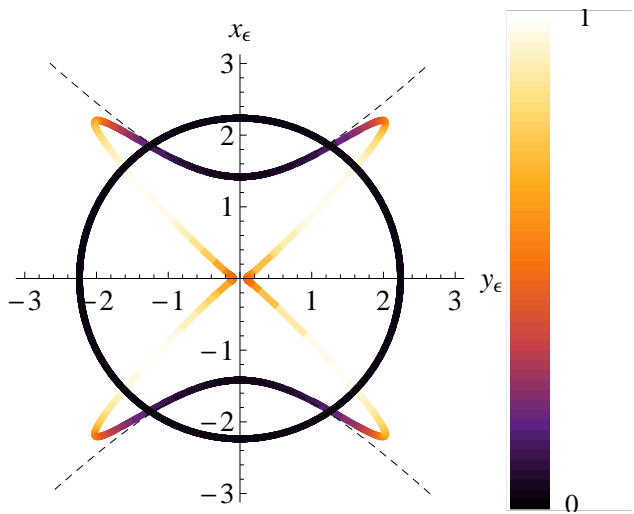}\\
  \caption{(color online) Isofrequency surface in $x_\epsilon$-$y_\epsilon$ 
  plane ($z_\epsilon=0$) showing conical intersection in the presence of 
  loss, with $\epsilon_1=-2+0.3 i,\epsilon_2=2+0.3 i$ and 
  $\epsilon_3=5+0.3 i$, similar to the bottom right panel of 
  Fig. \ref{fig:trans}. This is a polar plot of the real part of the 
  refractive index with direction, with color representing imaginary part 
  of the refractive index, i.e. the absorption. White represents solutions 
  with large absorption, and black those which are fully propagating. The 
  original intersection remains a mostly propagating solution. An 
  additional intersection appears which is mostly imaginary. The inclusion 
  of an imaginary component to the effective medium theory is enough to 
  prevent the dispersion surface becoming infinite. Dashed line shows 
  continuation of hyperbola in case of real $\epsilon$.
  \label{fig:loss}}
\end{figure}

Figure~\ref{fig:loss} shows the isofrequency surface in the
$x_\epsilon$-$y_\epsilon$ plane when each principal dielectric
constant has an imaginary part of $0.3$. This corresponds to an
isotropic absorption; anisotropic absorption does not qualitatively
change the results.  Note that an isotropic material with
$\epsilon=2+0.3i$ would have an imaginary refractive index of $\kappa=0.1$, 
meaning the decay length of the intensity $\lambda/(4\pi\kappa)$ would be 
less than a wavelength. Hence the imaginary part we are considering is 
small but not negigible. We see from Fig.\ \ref{fig:loss} that the crossings
identified in the absence of absorption remain, and are not destroyed
by the introduction of a complex permittivity. Furthermore the
wavevector at the crossings has a small imaginary component, relative
to its real component, meaning that these crossings correspond to
(mostly) propagating solutions with some absorption. The persistence
of intersections is ensured by the topological argument given
previously, as the absorption does not break the symmetry between left
and right circular polarizations.

We also note, from Fig.\ \ref{fig:loss}, that in the case of complex
dielectric constants the refractive index no longer goes to infinity:
the open hyperboloid becomes closed and finite. This is purely a
result of including losses, without leaving the effective medium
theory. The hyperboloid dispersion surface bends back at finite
$\bvec{k}$, intersecting the ellipsoidal surface again. This second
intersection has a large imaginary component, meaning that rays in
this direction will decay quickly. These new intersections also occur
in other directions of $\bvec{\eta}$, where they are also mainly
evanescent. As the imaginary component of $\epsilon$ is increased,
this finite hyperboloid shape will decrease in size, until the
mostly-real and mostly-imaginary intersections approach each other and
finally disappear. However mostly-imaginary intersections also appear
in the $x_\epsilon$-$z_\epsilon$ plane which remain for large
imaginary components, in keeping with our previous topological
argument.

\section{Diffraction}
\label{sec:diff}

A complete treatment of optics near the conical singuarities in a HMM
must allow for diffraction of the incident and refracted beams. Here
we develop such a treatment, and obtain formulas for the diffraction
patterns generated by arbitrary beams, incident on a biaxial HMM, with
wavevectors close to the optic axis. We follow the method of~\cite{Berry04},
in particular we use the angular spectrum
representation to calculate the contribution of each input ray to the
beam at a fixed propagation distance.  Describing beams propagating
close to the optic axis, which we will continue to label as the $x$
axis, the field at a position $x$ in the crystal consists of a sum of
plane wave components which pick up a phase on propagating
\begin{equation} \label{asr}
\begin{aligned}
E_{\mathrm{out}} = \iint \mathrm{d}k_y \mathrm{d}k_z\, E_{in}(k_y,k_z) 
\exp(i\left(k_y y+k_z z\right)) \\
\exp(i x \sqrt{k_T^2-k_y^2-k_z^2})
\end{aligned}
\end{equation}
where $E_{in}(k_y,k_z)$ is the two-dimensional Fourier transform of the 
input field in the plane $x=0$. However the magnitude of the total 
wavevector in the crystal $k_T$  is $n k_0$, with $n$ depending on the 
direction of the ray, i.e. on $k_y$ and $k_z$. We can express the 
refractive index given by Eq. \eqref{n2} in terms of the relative 
transverse momentum $\bvec{p}=k_\perp/k$, where $k=\sqrt{\epsilon_3} k_0$ 
is the magnitude of a wavevector lying directly along the optic axis. For 
small $\theta$ the transverse momenta are related to the angles defined 
in Fig. \ref{fig:coords} by $p_z= \theta \sin(\phi)$, $p_y =  \theta \cos(
\phi)$ and $p=\abs{\bvec{p}}=\theta$. The lowest order terms, linear in $p
$, lead to refraction into a simple cone which dominates the diffraction 
pattern. To reveal the fine structure we expand to second order giving 
\begin{equation}\begin{aligned}\label{n22} 
n^2 & \approx\epsilon_3-\epsilon_\delta (p_y\pm p)+
\left(\epsilon_\Delta p\pm \frac{\epsilon_\delta^2}{\epsilon_3} p_y\right)
\left(p\mp p_y\right)\\
&\equiv \epsilon_3 [1+\mu(p_y,p)]
\end{aligned}\end{equation}
where
\begin{equation}\label{edd} \epsilon_\Delta = 
\frac{\epsilon_3^2}{\epsilon_1 \epsilon_2}
\left(2\epsilon_3-\epsilon_1-\epsilon_2\right).
\end{equation}

Letting $k_T^2=n^2k_0^2=k^2\left(1+\mu(p_y,p)\right)$  we can expand the
square root in the final exponent of Eq. \eqref{asr}, again
to $\orderof(p^2)$ giving
\begin{equation}\label{exp} \begin{aligned}
\sqrt{k_T^2-k_{\perp}^2} &= \sqrt{n^2 k_0^2- k^2 p^2} \\
	&= k \sqrt{1+\mu(p,p_y)-p^2} \\
	&\approx k \left(1+\frac{1}{2}\mu(p,p_y)-
	\frac{1}{8}\mu(p,p_y)^2-\frac{1}{2}p^2\right)
	\end{aligned}\end{equation}
where we keep terms up to $\orderof(p^2)$ in $\mu^2$.

The integral Eq. \eqref{asr} with the approximation given in 
Eq. \eqref{exp} gives the paraxial approximation to the electric field at 
a plane $x> 0$, valid for small transverse momentum $p \ll 1$ or 
equivalently $k_ \perp\ll k$. The term in the exponent proportional to $x 
p_y$ leads to a skew away from the optic axis in the cone, as suggested 
by Fig. \ref{fig:cones}, which can be included in the definition of a 
new transverse coordinate which follows the center of the cone $\bvec{r}'_
\perp=\bvec{r}_ \perp+Ax\hat{e}_y$ such that $\bvec{p}\cdot\bvec{r}_\perp+
 Axp_y=\bvec{p} \cdot\left(\bvec{r}_\perp+A x \hat{e}_y\right)=\bvec{p}
\cdot\bvec{r}'_ \perp$. The remaining terms which depend on $p_y$ can not 
be absorbed in this way and lead to a non- circular asymmetry in the 
diffraction 
pattern.  
For simplicity we focus on the case of a circularly polarized, or
unpolarized, input beam. For each wavevector the two orthogonal
eigenpolarizations, given by Eq. \eqref{polarization}, will then be
present in equal amounts, and will not interfere with each other. We
consider a crystal of finite length $l<x$, so that the field
propagates a length $l$ through the crystal before propagating a
length $x-l$ in free space. Propagation beyond the crystal is described
by an identical integral to Eq. \eqref{asr} with the input field taken at
the plane $x=l$ and with $k_T=k_0$. The intensity at a point
$(x,\bvec{r}'_\perp)$ can then be written as the sum of the diffracted
intensities from each eigenpolarization,
\begin{equation} I=\abs{b_+}^2+\abs{b_-}^2.\label{eq:overalldiff}
\end{equation}
Expressing Eq. \eqref{asr} in terms of $p$ and using Eq. \eqref{exp} gives
\begin{equation}\label{bpm} \begin{aligned}
b_\pm(x,\bvec{r}'_\perp) = &\frac{k}{2\pi} e^{i k x}\iint \mathrm{d}^2p\, 
a(\bvec{p})\exp(ik\bvec{p}\cdot\bvec{r}'_\perp)\\
& \times\exp\left\{-ik p^2[\beta l+
\frac{1}{2}\sqrt{\epsilon_3}(x-l)]\right\}\\ 
&\times\exp(-ikl\alpha p_y^2) \\
&\times\exp[\pm iklp (\gamma+\delta p_y)]
\end{aligned} \end{equation} where $a(\bvec{p})$ is the Fourier
transform of the input field and $\alpha,\beta,\gamma$ and $\delta$
are all expressed in terms of $\epsilon_i$; the explicit forms are
given in the appendix. These parameters control the diffraction
patterns and have the following interpretations: $\beta$ is a
propagation constant, $\gamma$ is proportional to the angle of the
cone opening, and $\alpha$ and $\delta$ control the fine-structure of
the diffraction pattern leading to circular asymmetry.

\begin{figure}[htbp]
 \centering
 \includegraphics[width=\columnwidth]{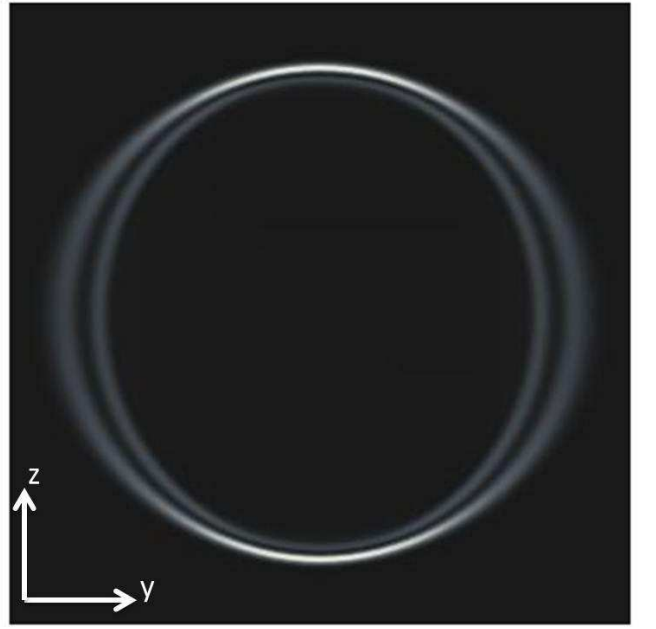}\\
  \caption{The intensity profile formed by conical diffraction of a 
Gaussian beam in a hyperbolic metamaterial, in the focal image plane 
(see text). The pattern is generated from the paraxial diffraction 
integral, Eq. \eqref{bpm}, with $\alpha l=10$ and $\delta=0$.   
\label{fig:diff}}
\end{figure}

As a specific application of the diffraction formulas,
Eqs.~\eqref{eq:overalldiff} and \eqref{bpm}), we show in
Fig. \ref{fig:diff} the conical diffraction pattern formed for a
Gaussian beam, $a(p)=k w^2 \exp(-k^2 p^2 w^2/2)$. The beam waist $w$
is taken as the unit length scale. The resulting intensity profile is
plotted in the focal image plane, $x=l-2\beta l/n_3$, where the
resulting ring structure is sharpest. This position corresponds to
the image of the input beam waist in an isotropic crystal of index $n_3$,
and the pattern here can be imaged with a lens if it occurs inside or
before the crystal~\cite{Berry04}. As $\alpha,\beta,\gamma,$ and
$\delta$ all appear multiplied by $l$ for propagation inside the
crystal, the length of the crystal is only important relative to the
overall scale of these parameters, e.g. a short, strongly diffracting
crystal will have the same effect as a long, weakly diffracting
one. The parameter $\gamma l$ is chosen to give a ring radius
$r_0\approx 50 w$ to ensure well developed rings while the other
parameters are $\alpha l =10,\delta l=0$. This choice allows us to
show the asymmetry of the beam on the same scale as the overall
conical refraction.

Like the positive $\epsilon$ case, the diffraction pattern consists of
two rings. In contrast to that case, however, the diffraction pattern
is not circularly symmetrical. The rings are broadened in the $y$
direction but remain tight in the z direction. This is in agreement
with Fig. \ref{fig:cones} which shows the cones moving apart in the
$y$ direction with increasing $p$. The diffraction pattern is bounded
approximately on the inside and the outside by the arcs of two
intersecting circles, also in agreement with the ray diagram. In
addition there is a dark ring. This is purely an effect of diffraction
and is not predicted by geometrical optics~\cite{Portigal69}. A
similar dark ring, known as the Pogendorff ring, also appears in the
conventional positive $\epsilon$ case.

\section{Discussion}
\label{sec:disc}

As discussed in the introduction, a key feature of our results is the
existence of linear intersections in the isofrequency surface in
HMMs. These resemble the Dirac points that are of great interest in
both condensed-matter physics and optics. It is therefore important to
consider the relation between these phenomena carefully.

The dispersion surfaces describing the propagation of light in a
biaxial material can be related to a bandstructure in two ways. The
most straightforward is to consider the full dispersion relation
$\omega(\bvec{k})$ of light, which is a surface in the four-dimensional
space of $\omega$ and $\bvec{k}$, and compare it with the corresponding
dispersion relation for electrons in a periodic lattice. In this case,
the isofrequency surfaces described here are directly equivalent to a
constant energy surface like the Fermi surface, and not directly to
the dispersion relation as usually plotted. Both are, of course,
cross-sections of the full dispersion relation in the four-dimensional
space of $\omega$ and $\bvec{k}$, but in different directions.

For electrons there are two spin states related by time reversal, so
that if time reversal symmetry is present
$\omega_+(\bvec{k})=\omega_-(\bvec{-k})$. If there is spatial inversion
symmetry then we furthermore have
$\omega_-(\bvec{k})=\omega_-(\bvec{-k})$. Hence if these two symmetries
are present there is only one, doubly degenerate, sheet to the Fermi
surface. This is a case of Kramer's degeneracy. If one of these
symmetries is broken then the spin up and spin down electrons can have
different Fermi surfaces which may have conical intersections
analogous to those described here, with the most common example being
ferromagnetism~\cite{Groot83}. 

For photons there are also two states, corresponding to the two
polarizations, but these are related not by time-reversal symmetry,
but by electro-magnetic duality. This symmetry is present if the
electric and magnetic fields can be interchanged. In most materials it
is broken, because $\epsilon\neq\mu$, and this allows full frequency
gaps to open, for example in a photonic crystal~\cite{Khanikaev13}. In
terms of the isofrequency surfaces the (usual) breaking of this
symmetry lifts the polarization degeneracy for most directions,
leaving only the isolated point singularities described here. 

There is, however, a less immediate but stronger connection between
conical singularities and Dirac points, based on the well-known
equivalence between the Schrodinger equation in $2+1\mathrm{D}$ and
the paraxial Helmholtz equation. To demonstrate this connection in the
present case, we construct the paraxial Helmholtz equation describing
conical diffraction in a biaxial HMM. We begin by writing the electric
field as a plane wave times a slowly varying envelope function
\begin{equation}
  E(\bvec{r})=A(\bvec{r}) \exp(i k x), \end{equation}
where $A(\bvec{r})$ varies slowly with $x$. 
The diffracted field given 
by Eq. \eqref{bpm} can be expressed as the two-dimensional transverse
input field 
evolving in the $x$-direction as
\begin{equation}
E(\bvec{r}_\perp,x) =\exp(-ik\int_0^x\mathrm{d}x'\,H(p,x'))E(\bvec{r}_\perp,0)
\end{equation}
where for conical diffraction in a HMM we find that the Hamiltonian
is
\begin{equation}\label{eq:paraxh}
H=\alpha p_y^2 + \beta p^2 +(\gamma+\delta p_y) \bvec{s}\cdot\bvec{p}
\end{equation}
for $x<l$, and is the free Hamiltonian $p^2/2$ for $x>l$. Here
$\bvec{s}=\left\{\sigma_3,\sigma_1\right\}$ is a vector of Pauli
matrices in a Cartesian basis and $\bvec{p}$ is formally represented by
$-i\nabla_\perp/k$.  The envelope function, thus, obeys the paraxial
Helmholtz equation, which takes the form
\begin{equation} H A = \frac{i}{k}\frac{\partial A}{\partial
    x}. \label{eq:parahelm}\end{equation} Since this is equivalent to
the Schr\"odinger equation~\cite{Dragoman04}, the propagation with $x$
of the two dimensional transverse beam is equivalent to the evolution
with time of the wavefunction for a spin-$1/2$ particle. The
birefringence of a biaxial material appears as a spin-orbit coupling,
whose explicit form, close to the optic axis for a HMM, can be seen in
Eq. \eqref{eq:paraxh}. This form, with different definitions of the
constants, also applies to a conventional biaxial material, but in
that case the anisotropic terms proportional to $\alpha$ and $\delta$
are negligible and can be dropped~\cite{Berry04}.

Since light (of a fixed frequency) propagates in space according to
Eq. \eqref{eq:parahelm}, with $x$ playing the role of time, the
propagation constant $k_x$ can be interpreted as the energy. The
isofrequency surfaces can thus be seen as a dispersion relation,
giving the propagation constant as a function of the two transverse
momenta $k_y,k_z$. The point intersections in the isofrequency
surfaces then correspond to Dirac points for two-dimensional
electrons; specifically, the point intersections discussed here are
the Dirac points of the Hamiltonian, Eq. \eqref{eq:paraxh}.

Dirac points in two dimensional materials have been of interest for their 
role in topological insulators and topologically protected edge states~
\cite{Haldane88,Kane05}. In a hexagonal lattice such as graphene, subject 
to time reversal symmetry and spatial inversion symmetry, the electronic 
band stucture must contain Dirac points. These degeneracies can be lifted 
by breaking spatial inversion symmetry, leading to a trivial insulator, 
or by breaking time reversal symmetry, leading to a topological insulator~
\cite{Hasan10}. Hence, work on topological effects in photonic systems 
has focused on Dirac points, primarily in the full frequency dispersion 
$\omega(\bvec{k})$~\cite{Raghu08,Ochiai09,Khanikaev13}. More recently 
however attention has shifted to the analogous Dirac, or conical, 
intersections in the paraxial propagation constant surface~\cite{
Rechtsman13b,Gao14,Rechtsman13}. Understanding the effects of different 
symmetries on these two dispersion surfaces could therefore help progress 
towards topologically protected photonic systems.

\section{Conclusions}

\label{sec:conc}

These results illustrate the unique singularities found in hyperbolic 
metamaterials when all three indices are allowed to vary independently. 
By examining the full dispersion surface of a general, biaxial, 
hyperbolic metamaterial, we have identified conical singularities at 
which the refraction direction is not defined. We have found the 
approximate dispersion surface and the refracted Poynting vector for a 
ray traveling close to the axis of these singularities. We have shown 
that this leads to a new form of refraction which does not appear in the 
usual uniaxial HMMs and is topologically and quantitatively different 
from the phenomenon of conical refraction which occurs in ordinary 
biaxial materials. These propagating solutions remain when a small 
imaginary component is included, leading to a small amount of absorption, 
with additional mostly evanescent singular solutions also appearing. We 
have also calculated the diffraction pattern for a beam traveling through 
such a material. We have found that the diffracted beam is generally not 
circularly symmetric and that, similar to the positive $\epsilon$ case, a 
dark ring appears where ray optics predicts the largest intensity.

\begin{acknowledgments}
This work was supported by Science Foundation Ireland grant SIRG I/1592 and by
the Higher Education Authority under PRTLI funding cycle 5. The authors wish
to thank Prof. J. G. Lunney for useful discussions.
\end{acknowledgments}

\appendix*
\section{}
We provide the parameters used in the diffraction theory in terms of the 
dielectric constants of the material:

\begin{equation}\begin{aligned}
\alpha &= \frac{\epsilon_\delta}{8}-\frac{\epsilon_\Delta^2}{2\epsilon_3} \\
\beta &= \frac{1}{2}\left(\epsilon_\Delta-1\right)
+\frac{1}{8}\epsilon_\delta \\
\gamma &=  \frac{1}{2}\epsilon_\delta \\
\delta &= \frac{\epsilon_\delta^2}{2 
\epsilon_3}+\frac{\epsilon_\delta}{4}-\frac{\epsilon_\Delta}{2},
\end{aligned}\end{equation}

recalling from Eqs. \eqref{ed} and \eqref{edd} that

\begin{equation}\begin{aligned}
\epsilon_\delta &= \epsilon_3\sqrt{\frac{\left(\epsilon_3-\epsilon_1\right) 
\left(\epsilon
_2-\epsilon_3\right)}{\epsilon_1 \epsilon_2}} \\
\epsilon_\Delta &= \frac{\epsilon_3^2}{\epsilon_1 
\epsilon_2}\left(2\epsilon_3-\epsilon_1-\epsilon_2\right).
\end{aligned}\end{equation}


\begin{thebibliography}{49}%
\makeatletter
\providecommand \@ifxundefined [1]{%
 \@ifx{#1\undefined}
}%
\providecommand \@ifnum [1]{%
 \ifnum #1\expandafter \@firstoftwo
 \else \expandafter \@secondoftwo
 \fi
}%
\providecommand \@ifx [1]{%
 \ifx #1\expandafter \@firstoftwo
 \else \expandafter \@secondoftwo
 \fi
}%
\providecommand \natexlab [1]{#1}%
\providecommand \enquote  [1]{``#1''}%
\providecommand \bibnamefont  [1]{#1}%
\providecommand \bibfnamefont [1]{#1}%
\providecommand \citenamefont [1]{#1}%
\providecommand \href@noop [0]{\@secondoftwo}%
\providecommand \href [0]{\begingroup \@sanitize@url \@href}%
\providecommand \@href[1]{\@@startlink{#1}\@@href}%
\providecommand \@@href[1]{\endgroup#1\@@endlink}%
\providecommand \@sanitize@url [0]{\catcode `\\12\catcode `\$12\catcode
  `\&12\catcode `\#12\catcode `\^12\catcode `\_12\catcode `\%12\relax}%
\providecommand \@@startlink[1]{}%
\providecommand \@@endlink[0]{}%
\providecommand \url  [0]{\begingroup\@sanitize@url \@url }%
\providecommand \@url [1]{\endgroup\@href {#1}{\urlprefix }}%
\providecommand \urlprefix  [0]{URL }%
\providecommand \Eprint [0]{\href }%
\providecommand \doibase [0]{http://dx.doi.org/}%
\providecommand \selectlanguage [0]{\@gobble}%
\providecommand \bibinfo  [0]{\@secondoftwo}%
\providecommand \bibfield  [0]{\@secondoftwo}%
\providecommand \translation [1]{[#1]}%
\providecommand \BibitemOpen [0]{}%
\providecommand \bibitemStop [0]{}%
\providecommand \bibitemNoStop [0]{.\EOS\space}%
\providecommand \EOS [0]{\spacefactor3000\relax}%
\providecommand \BibitemShut  [1]{\csname bibitem#1\endcsname}%
\let\auto@bib@innerbib\@empty
\bibitem [{\citenamefont {Cortes}\ \emph {et~al.}(2012)\citenamefont {Cortes},
  \citenamefont {Newman}, \citenamefont {Molesky},\ and\ \citenamefont
  {Jacob}}]{Cortes12}%
  \BibitemOpen
  \bibfield  {author} {\bibinfo {author} {\bibfnamefont {C.~L.}\ \bibnamefont
  {Cortes}}, \bibinfo {author} {\bibfnamefont {W.}~\bibnamefont {Newman}},
  \bibinfo {author} {\bibfnamefont {S.}~\bibnamefont {Molesky}}, \ and\
  \bibinfo {author} {\bibfnamefont {Z.}~\bibnamefont {Jacob}},\ }\href
  {http://stacks.iop.org/2040-8986/14/i=6/a=063001} {\bibfield  {journal}
  {\bibinfo  {journal} {Journal of Optics}\ }\textbf {\bibinfo {volume} {14}},\
  \bibinfo {pages} {063001} (\bibinfo {year} {2012})}\BibitemShut {NoStop}%
\bibitem [{\citenamefont {Podolskiy}\ and\ \citenamefont
  {Narimanov}(2005)}]{Podolskiy05}%
  \BibitemOpen
  \bibfield  {author} {\bibinfo {author} {\bibfnamefont {V.~A.}\ \bibnamefont
  {Podolskiy}}\ and\ \bibinfo {author} {\bibfnamefont {E.~E.}\ \bibnamefont
  {Narimanov}},\ }\href {\doibase 10.1103/PhysRevB.71.201101} {\bibfield
  {journal} {\bibinfo  {journal} {Phys. Rev. B}\ }\textbf {\bibinfo {volume}
  {71}},\ \bibinfo {pages} {201101} (\bibinfo {year} {2005})}\BibitemShut
  {NoStop}%
\bibitem [{\citenamefont {Kidwai}\ \emph {et~al.}(2012)\citenamefont {Kidwai},
  \citenamefont {Zhukovsky},\ and\ \citenamefont {Sipe}}]{Kidwai12}%
  \BibitemOpen
  \bibfield  {author} {\bibinfo {author} {\bibfnamefont {O.}~\bibnamefont
  {Kidwai}}, \bibinfo {author} {\bibfnamefont {S.~V.}\ \bibnamefont
  {Zhukovsky}}, \ and\ \bibinfo {author} {\bibfnamefont {J.~E.}\ \bibnamefont
  {Sipe}},\ }\href {\doibase 10.1103/PhysRevA.85.053842} {\bibfield  {journal}
  {\bibinfo  {journal} {Phys. Rev. A}\ }\textbf {\bibinfo {volume} {85}},\
  \bibinfo {pages} {053842} (\bibinfo {year} {2012})}\BibitemShut {NoStop}%
\bibitem [{\citenamefont {Jacob}\ \emph {et~al.}(2012)\citenamefont {Jacob},
  \citenamefont {Smolyaninov},\ and\ \citenamefont {Narimanov}}]{Jacob12}%
  \BibitemOpen
  \bibfield  {author} {\bibinfo {author} {\bibfnamefont {Z.}~\bibnamefont
  {Jacob}}, \bibinfo {author} {\bibfnamefont {I.~I.}\ \bibnamefont
  {Smolyaninov}}, \ and\ \bibinfo {author} {\bibfnamefont {E.~E.}\ \bibnamefont
  {Narimanov}},\ }\href {\doibase 10.1063/1.4710548} {\bibfield  {journal}
  {\bibinfo  {journal} {Applied Physics Letters}\ }\textbf {\bibinfo {volume}
  {100}},\ \bibinfo {eid} {181105} (\bibinfo {year} {2012})}\BibitemShut
  {NoStop}%
\bibitem [{\citenamefont {Yang}\ \emph {et~al.}(2012)\citenamefont {Yang},
  \citenamefont {Yao}, \citenamefont {Rho}, \citenamefont {Yin},\ and\
  \citenamefont {Zhang}}]{Yang12}%
  \BibitemOpen
  \bibfield  {author} {\bibinfo {author} {\bibfnamefont {X.}~\bibnamefont
  {Yang}}, \bibinfo {author} {\bibfnamefont {J.}~\bibnamefont {Yao}}, \bibinfo
  {author} {\bibfnamefont {J.}~\bibnamefont {Rho}}, \bibinfo {author}
  {\bibfnamefont {X.}~\bibnamefont {Yin}}, \ and\ \bibinfo {author}
  {\bibfnamefont {X.}~\bibnamefont {Zhang}},\ }\href@noop {} {\bibfield
  {journal} {\bibinfo  {journal} {Nat Photon}\ }\textbf {\bibinfo {volume}
  {6}},\ \bibinfo {pages} {450} (\bibinfo {year} {2012})}\BibitemShut {NoStop}%
\bibitem [{\citenamefont {Smith}\ \emph {et~al.}(2004)\citenamefont {Smith},
  \citenamefont {Kolinko},\ and\ \citenamefont {Schurig}}]{Smith04}%
  \BibitemOpen
  \bibfield  {author} {\bibinfo {author} {\bibfnamefont {D.~R.}\ \bibnamefont
  {Smith}}, \bibinfo {author} {\bibfnamefont {P.}~\bibnamefont {Kolinko}}, \
  and\ \bibinfo {author} {\bibfnamefont {D.}~\bibnamefont {Schurig}},\ }\href
  {\doibase 10.1364/JOSAB.21.001032} {\bibfield  {journal} {\bibinfo  {journal}
  {J. Opt. Soc. Am. B}\ }\textbf {\bibinfo {volume} {21}},\ \bibinfo {pages}
  {1032} (\bibinfo {year} {2004})}\BibitemShut {NoStop}%
\bibitem [{\citenamefont {Jacob}\ \emph {et~al.}(2006)\citenamefont {Jacob},
  \citenamefont {Alekseyev},\ and\ \citenamefont {Narimanov}}]{Jacob06}%
  \BibitemOpen
  \bibfield  {author} {\bibinfo {author} {\bibfnamefont {Z.}~\bibnamefont
  {Jacob}}, \bibinfo {author} {\bibfnamefont {L.~V.}\ \bibnamefont
  {Alekseyev}}, \ and\ \bibinfo {author} {\bibfnamefont {E.}~\bibnamefont
  {Narimanov}},\ }\href {\doibase 10.1364/OE.14.008247} {\bibfield  {journal}
  {\bibinfo  {journal} {Opt. Express}\ }\textbf {\bibinfo {volume} {14}},\
  \bibinfo {pages} {8247} (\bibinfo {year} {2006})}\BibitemShut {NoStop}%
\bibitem [{\citenamefont {Liu}\ \emph {et~al.}(2007)\citenamefont {Liu},
  \citenamefont {Lee}, \citenamefont {Xiong}, \citenamefont {Sun},\ and\
  \citenamefont {Zhang}}]{Liu07}%
  \BibitemOpen
  \bibfield  {author} {\bibinfo {author} {\bibfnamefont {Z.}~\bibnamefont
  {Liu}}, \bibinfo {author} {\bibfnamefont {H.}~\bibnamefont {Lee}}, \bibinfo
  {author} {\bibfnamefont {Y.}~\bibnamefont {Xiong}}, \bibinfo {author}
  {\bibfnamefont {C.}~\bibnamefont {Sun}}, \ and\ \bibinfo {author}
  {\bibfnamefont {X.}~\bibnamefont {Zhang}},\ }\href {\doibase
  10.1126/science.1137368} {\bibfield  {journal} {\bibinfo  {journal}
  {Science}\ }\textbf {\bibinfo {volume} {315}},\ \bibinfo {pages} {1686}
  (\bibinfo {year} {2007})}\BibitemShut {NoStop}%
\bibitem [{\citenamefont {Kabashin}\ \emph {et~al.}(2009)\citenamefont
  {Kabashin}, \citenamefont {Evans}, \citenamefont {Pastkovsky}, \citenamefont
  {Hendren}, \citenamefont {Wurtz}, \citenamefont {Atkinson}, \citenamefont
  {Pollard}, \citenamefont {Podolskiy},\ and\ \citenamefont
  {Zayats}}]{Kabashin09}%
  \BibitemOpen
  \bibfield  {author} {\bibinfo {author} {\bibfnamefont {A.~V.}\ \bibnamefont
  {Kabashin}}, \bibinfo {author} {\bibfnamefont {P.}~\bibnamefont {Evans}},
  \bibinfo {author} {\bibfnamefont {S.}~\bibnamefont {Pastkovsky}}, \bibinfo
  {author} {\bibfnamefont {W.}~\bibnamefont {Hendren}}, \bibinfo {author}
  {\bibfnamefont {G.~A.}\ \bibnamefont {Wurtz}}, \bibinfo {author}
  {\bibfnamefont {R.}~\bibnamefont {Atkinson}}, \bibinfo {author}
  {\bibfnamefont {R.}~\bibnamefont {Pollard}}, \bibinfo {author} {\bibfnamefont
  {V.~A.}\ \bibnamefont {Podolskiy}}, \ and\ \bibinfo {author} {\bibfnamefont
  {A.~V.}\ \bibnamefont {Zayats}},\ }\href@noop {} {\bibfield  {journal}
  {\bibinfo  {journal} {Nat Mater}\ }\textbf {\bibinfo {volume} {8}},\ \bibinfo
  {pages} {867} (\bibinfo {year} {2009})}\BibitemShut {NoStop}%
\bibitem [{\citenamefont {Govyadinov}\ and\ \citenamefont
  {Podolskiy}(2006)}]{Govyadinov06}%
  \BibitemOpen
  \bibfield  {author} {\bibinfo {author} {\bibfnamefont {A.~A.}\ \bibnamefont
  {Govyadinov}}\ and\ \bibinfo {author} {\bibfnamefont {V.~A.}\ \bibnamefont
  {Podolskiy}},\ }\href {\doibase 10.1103/PhysRevB.73.155108} {\bibfield
  {journal} {\bibinfo  {journal} {Phys. Rev. B}\ }\textbf {\bibinfo {volume}
  {73}},\ \bibinfo {pages} {155108} (\bibinfo {year} {2006})}\BibitemShut
  {NoStop}%
\bibitem [{\citenamefont {He}\ \emph {et~al.}(2012)\citenamefont {He},
  \citenamefont {He}, \citenamefont {Gao},\ and\ \citenamefont {Yang}}]{He12}%
  \BibitemOpen
  \bibfield  {author} {\bibinfo {author} {\bibfnamefont {Y.}~\bibnamefont
  {He}}, \bibinfo {author} {\bibfnamefont {S.}~\bibnamefont {He}}, \bibinfo
  {author} {\bibfnamefont {J.}~\bibnamefont {Gao}}, \ and\ \bibinfo {author}
  {\bibfnamefont {X.}~\bibnamefont {Yang}},\ }\href {\doibase
  10.1364/JOSAB.29.002559} {\bibfield  {journal} {\bibinfo  {journal} {J. Opt.
  Soc. Am. B}\ }\textbf {\bibinfo {volume} {29}},\ \bibinfo {pages} {2559}
  (\bibinfo {year} {2012})}\BibitemShut {NoStop}%
\bibitem [{\citenamefont {Wurtz}\ \emph {et~al.}(2011)\citenamefont {Wurtz},
  \citenamefont {PollardR}, \citenamefont {HendrenW}, \citenamefont
  {Wiederrecht}, \citenamefont {Gosztola}, \citenamefont {Podolskiy},\ and\
  \citenamefont {Zayats}}]{Wurtz11}%
  \BibitemOpen
  \bibfield  {author} {\bibinfo {author} {\bibfnamefont {G.~A.}\ \bibnamefont
  {Wurtz}}, \bibinfo {author} {\bibnamefont {PollardR}}, \bibinfo {author}
  {\bibnamefont {HendrenW}}, \bibinfo {author} {\bibfnamefont {G.~P.}\
  \bibnamefont {Wiederrecht}}, \bibinfo {author} {\bibfnamefont {D.~J.}\
  \bibnamefont {Gosztola}}, \bibinfo {author} {\bibfnamefont {V.~A.}\
  \bibnamefont {Podolskiy}}, \ and\ \bibinfo {author} {\bibfnamefont {A.~V.}\
  \bibnamefont {Zayats}},\ }\href@noop {} {\bibfield  {journal} {\bibinfo
  {journal} {Nat Nano}\ }\textbf {\bibinfo {volume} {6}},\ \bibinfo {pages}
  {107} (\bibinfo {year} {2011})}\BibitemShut {NoStop}%
\bibitem [{\citenamefont {Krishnamoorthy}\ \emph {et~al.}(2012)\citenamefont
  {Krishnamoorthy}, \citenamefont {Jacob}, \citenamefont {Narimanov},
  \citenamefont {Kretzschmar},\ and\ \citenamefont {Menon}}]{Krishna12}%
  \BibitemOpen
  \bibfield  {author} {\bibinfo {author} {\bibfnamefont {H.~N.~S.}\
  \bibnamefont {Krishnamoorthy}}, \bibinfo {author} {\bibfnamefont
  {Z.}~\bibnamefont {Jacob}}, \bibinfo {author} {\bibfnamefont
  {E.}~\bibnamefont {Narimanov}}, \bibinfo {author} {\bibfnamefont
  {I.}~\bibnamefont {Kretzschmar}}, \ and\ \bibinfo {author} {\bibfnamefont
  {V.~M.}\ \bibnamefont {Menon}},\ }\href {\doibase 10.1126/science.1219171}
  {\bibfield  {journal} {\bibinfo  {journal} {Science}\ }\textbf {\bibinfo
  {volume} {336}},\ \bibinfo {pages} {205} (\bibinfo {year}
  {2012})}\BibitemShut {NoStop}%
\bibitem [{\citenamefont {Elser}\ \emph {et~al.}(2006)\citenamefont {Elser},
  \citenamefont {Wangberg}, \citenamefont {Podolskiy},\ and\ \citenamefont
  {Narimanov}}]{Elser06}%
  \BibitemOpen
  \bibfield  {author} {\bibinfo {author} {\bibfnamefont {J.}~\bibnamefont
  {Elser}}, \bibinfo {author} {\bibfnamefont {R.}~\bibnamefont {Wangberg}},
  \bibinfo {author} {\bibfnamefont {V.~A.}\ \bibnamefont {Podolskiy}}, \ and\
  \bibinfo {author} {\bibfnamefont {E.~E.}\ \bibnamefont {Narimanov}},\ }\href
  {\doibase http://dx.doi.org/10.1063/1.2422893} {\bibfield  {journal}
  {\bibinfo  {journal} {Applied Physics Letters}\ }\textbf {\bibinfo {volume}
  {89}},\ \bibinfo {eid} {261102} (\bibinfo {year} {2006})}\BibitemShut
  {NoStop}%
\bibitem [{\citenamefont {Sun}\ \emph {et~al.}(2013{\natexlab{a}})\citenamefont
  {Sun}, \citenamefont {Zeng},\ and\ \citenamefont {Litchinitser}}]{Sun13}%
  \BibitemOpen
  \bibfield  {author} {\bibinfo {author} {\bibfnamefont {J.}~\bibnamefont
  {Sun}}, \bibinfo {author} {\bibfnamefont {J.}~\bibnamefont {Zeng}}, \ and\
  \bibinfo {author} {\bibfnamefont {N.~M.}\ \bibnamefont {Litchinitser}},\
  }\href {\doibase 10.1364/OE.21.014975} {\bibfield  {journal} {\bibinfo
  {journal} {Opt. Express}\ }\textbf {\bibinfo {volume} {21}},\ \bibinfo
  {pages} {14975} (\bibinfo {year} {2013}{\natexlab{a}})}\BibitemShut {NoStop}%
\bibitem [{\citenamefont {Berry}\ and\ \citenamefont {Dennis}(2003)}]{Berry03}%
  \BibitemOpen
  \bibfield  {author} {\bibinfo {author} {\bibfnamefont {M.~V.}\ \bibnamefont
  {Berry}}\ and\ \bibinfo {author} {\bibfnamefont {M.~R.}\ \bibnamefont
  {Dennis}},\ }\href {\doibase 10.1098/rspa.2003.1155} {\bibfield  {journal}
  {\bibinfo  {journal} {Proceedings of the Royal Society of London. Series A:
  Mathematical, Physical and Engineering Sciences}\ }\textbf {\bibinfo {volume}
  {459}},\ \bibinfo {pages} {1261} (\bibinfo {year} {2003})}\BibitemShut
  {NoStop}%
\bibitem [{\citenamefont {Berry}(2004)}]{Berry04}%
  \BibitemOpen
  \bibfield  {author} {\bibinfo {author} {\bibfnamefont {M.~V.}\ \bibnamefont
  {Berry}},\ }\href {http://stacks.iop.org/1464-4258/6/i=4/a=001} {\bibfield
  {journal} {\bibinfo  {journal} {Journal of Optics A: Pure and Applied
  Optics}\ }\textbf {\bibinfo {volume} {6}},\ \bibinfo {pages} {289} (\bibinfo
  {year} {2004})}\BibitemShut {NoStop}%
\bibitem [{\citenamefont {Portigal}\ and\ \citenamefont
  {Burstein}(1969)}]{Portigal69}%
  \BibitemOpen
  \bibfield  {author} {\bibinfo {author} {\bibfnamefont {D.~L.}\ \bibnamefont
  {Portigal}}\ and\ \bibinfo {author} {\bibfnamefont {E.}~\bibnamefont
  {Burstein}},\ }\href {\doibase 10.1364/JOSA.59.001567} {\bibfield  {journal}
  {\bibinfo  {journal} {J. Opt. Soc. Am.}\ }\textbf {\bibinfo {volume} {59}},\
  \bibinfo {pages} {1567} (\bibinfo {year} {1969})}\BibitemShut {NoStop}%
\bibitem [{\citenamefont {Wallace}(1947)}]{Wallace47}%
  \BibitemOpen
  \bibfield  {author} {\bibinfo {author} {\bibfnamefont {P.~R.}\ \bibnamefont
  {Wallace}},\ }\href {\doibase 10.1103/PhysRev.71.622} {\bibfield  {journal}
  {\bibinfo  {journal} {Phys. Rev.}\ }\textbf {\bibinfo {volume} {71}},\
  \bibinfo {pages} {622} (\bibinfo {year} {1947})}\BibitemShut {NoStop}%
\bibitem [{\citenamefont {Novoselov}\ \emph {et~al.}(2004)\citenamefont
  {Novoselov}, \citenamefont {Geim}, \citenamefont {Morozov}, \citenamefont
  {Jiang}, \citenamefont {Zhang}, \citenamefont {Dubonos}, \citenamefont
  {Grigorieva},\ and\ \citenamefont {Firsov}}]{Novoselov04}%
  \BibitemOpen
  \bibfield  {author} {\bibinfo {author} {\bibfnamefont {K.~S.}\ \bibnamefont
  {Novoselov}}, \bibinfo {author} {\bibfnamefont {A.~K.}\ \bibnamefont {Geim}},
  \bibinfo {author} {\bibfnamefont {S.~V.}\ \bibnamefont {Morozov}}, \bibinfo
  {author} {\bibfnamefont {D.}~\bibnamefont {Jiang}}, \bibinfo {author}
  {\bibfnamefont {Y.}~\bibnamefont {Zhang}}, \bibinfo {author} {\bibfnamefont
  {S.~V.}\ \bibnamefont {Dubonos}}, \bibinfo {author} {\bibfnamefont {I.~V.}\
  \bibnamefont {Grigorieva}}, \ and\ \bibinfo {author} {\bibfnamefont {A.~A.}\
  \bibnamefont {Firsov}},\ }\href {\doibase 10.1126/science.1102896} {\bibfield
   {journal} {\bibinfo  {journal} {Science}\ }\textbf {\bibinfo {volume}
  {306}},\ \bibinfo {pages} {666} (\bibinfo {year} {2004})}\BibitemShut
  {NoStop}%
\bibitem [{\citenamefont {Castro~Neto}\ \emph {et~al.}(2009)\citenamefont
  {Castro~Neto}, \citenamefont {Guinea}, \citenamefont {Peres}, \citenamefont
  {Novoselov},\ and\ \citenamefont {Geim}}]{Castro09}%
  \BibitemOpen
  \bibfield  {author} {\bibinfo {author} {\bibfnamefont {A.~H.}\ \bibnamefont
  {Castro~Neto}}, \bibinfo {author} {\bibfnamefont {F.}~\bibnamefont {Guinea}},
  \bibinfo {author} {\bibfnamefont {N.~M.~R.}\ \bibnamefont {Peres}}, \bibinfo
  {author} {\bibfnamefont {K.~S.}\ \bibnamefont {Novoselov}}, \ and\ \bibinfo
  {author} {\bibfnamefont {A.~K.}\ \bibnamefont {Geim}},\ }\href {\doibase
  10.1103/RevModPhys.81.109} {\bibfield  {journal} {\bibinfo  {journal} {Rev.
  Mod. Phys.}\ }\textbf {\bibinfo {volume} {81}},\ \bibinfo {pages} {109}
  (\bibinfo {year} {2009})}\BibitemShut {NoStop}%
\bibitem [{\citenamefont {Geim}\ and\ \citenamefont
  {Novoselov}(2007)}]{Geim07}%
  \BibitemOpen
  \bibfield  {author} {\bibinfo {author} {\bibfnamefont {A.~K.}\ \bibnamefont
  {Geim}}\ and\ \bibinfo {author} {\bibfnamefont {K.~S.}\ \bibnamefont
  {Novoselov}},\ }\href {http://dx.doi.org/10.1038/nmat1849} {\bibfield
  {journal} {\bibinfo  {journal} {Nat Mater}\ }\textbf {\bibinfo {volume}
  {6}},\ \bibinfo {pages} {183} (\bibinfo {year} {2007})}\BibitemShut {NoStop}%
\bibitem [{\citenamefont {{Cooper}}\ \emph {et~al.}(2011)\citenamefont
  {{Cooper}}, \citenamefont {{D'Anjou}}, \citenamefont {{Ghattamaneni}},
  \citenamefont {{Harack}}, \citenamefont {{Hilke}}, \citenamefont {{Horth}},
  \citenamefont {{Majlis}}, \citenamefont {{Massicotte}}, \citenamefont
  {{Vandsburger}}, \citenamefont {{Whiteway}},\ and\ \citenamefont
  {{Yu}}}]{Cooper11}%
  \BibitemOpen
  \bibfield  {author} {\bibinfo {author} {\bibfnamefont {D.~R.}\ \bibnamefont
  {{Cooper}}}, \bibinfo {author} {\bibfnamefont {B.}~\bibnamefont {{D'Anjou}}},
  \bibinfo {author} {\bibfnamefont {N.}~\bibnamefont {{Ghattamaneni}}},
  \bibinfo {author} {\bibfnamefont {B.}~\bibnamefont {{Harack}}}, \bibinfo
  {author} {\bibfnamefont {M.}~\bibnamefont {{Hilke}}}, \bibinfo {author}
  {\bibfnamefont {A.}~\bibnamefont {{Horth}}}, \bibinfo {author} {\bibfnamefont
  {N.}~\bibnamefont {{Majlis}}}, \bibinfo {author} {\bibfnamefont
  {M.}~\bibnamefont {{Massicotte}}}, \bibinfo {author} {\bibfnamefont
  {L.}~\bibnamefont {{Vandsburger}}}, \bibinfo {author} {\bibfnamefont
  {E.}~\bibnamefont {{Whiteway}}}, \ and\ \bibinfo {author} {\bibfnamefont
  {V.}~\bibnamefont {{Yu}}},\ }\href@noop {} {\bibfield  {journal} {\bibinfo
  {journal} {ArXiv e-prints}\ } (\bibinfo {year} {2011})},\ \Eprint
  {http://arxiv.org/abs/1110.6557} {arXiv:1110.6557 [cond-mat.mes-hall]}
  \BibitemShut {NoStop}%
\bibitem [{\citenamefont {Das~Sarma}\ \emph {et~al.}(2011)\citenamefont
  {Das~Sarma}, \citenamefont {Adam}, \citenamefont {Hwang},\ and\ \citenamefont
  {Rossi}}]{Sarma11}%
  \BibitemOpen
  \bibfield  {author} {\bibinfo {author} {\bibfnamefont {S.}~\bibnamefont
  {Das~Sarma}}, \bibinfo {author} {\bibfnamefont {S.}~\bibnamefont {Adam}},
  \bibinfo {author} {\bibfnamefont {E.~H.}\ \bibnamefont {Hwang}}, \ and\
  \bibinfo {author} {\bibfnamefont {E.}~\bibnamefont {Rossi}},\ }\href
  {\doibase 10.1103/RevModPhys.83.407} {\bibfield  {journal} {\bibinfo
  {journal} {Rev. Mod. Phys.}\ }\textbf {\bibinfo {volume} {83}},\ \bibinfo
  {pages} {407} (\bibinfo {year} {2011})}\BibitemShut {NoStop}%
\bibitem [{\citenamefont {Novoselov}\ \emph {et~al.}(2005)\citenamefont
  {Novoselov}, \citenamefont {Geim}, \citenamefont {Morozov}, \citenamefont
  {Jiang}, \citenamefont {Katsnelson}, \citenamefont {Grigorieva},
  \citenamefont {Dubonos},\ and\ \citenamefont {Firsov}}]{Novoselov05}%
  \BibitemOpen
  \bibfield  {author} {\bibinfo {author} {\bibfnamefont {K.~S.}\ \bibnamefont
  {Novoselov}}, \bibinfo {author} {\bibfnamefont {A.~K.}\ \bibnamefont {Geim}},
  \bibinfo {author} {\bibfnamefont {S.~V.}\ \bibnamefont {Morozov}}, \bibinfo
  {author} {\bibfnamefont {D.}~\bibnamefont {Jiang}}, \bibinfo {author}
  {\bibfnamefont {M.~I.}\ \bibnamefont {Katsnelson}}, \bibinfo {author}
  {\bibfnamefont {I.~V.}\ \bibnamefont {Grigorieva}}, \bibinfo {author}
  {\bibfnamefont {S.~V.}\ \bibnamefont {Dubonos}}, \ and\ \bibinfo {author}
  {\bibfnamefont {A.~A.}\ \bibnamefont {Firsov}},\ }\href@noop {} {\bibfield
  {journal} {\bibinfo  {journal} {Nature}\ }\textbf {\bibinfo {volume} {438}},\
  \bibinfo {pages} {197} (\bibinfo {year} {2005})}\BibitemShut {NoStop}%
\bibitem [{\citenamefont {Gusynin}\ and\ \citenamefont
  {Sharapov}(2005)}]{Gusynin05}%
  \BibitemOpen
  \bibfield  {author} {\bibinfo {author} {\bibfnamefont {V.~P.}\ \bibnamefont
  {Gusynin}}\ and\ \bibinfo {author} {\bibfnamefont {S.~G.}\ \bibnamefont
  {Sharapov}},\ }\href {\doibase 10.1103/PhysRevLett.95.146801} {\bibfield
  {journal} {\bibinfo  {journal} {Phys. Rev. Lett.}\ }\textbf {\bibinfo
  {volume} {95}},\ \bibinfo {pages} {146801} (\bibinfo {year}
  {2005})}\BibitemShut {NoStop}%
\bibitem [{\citenamefont {Zhang}\ \emph {et~al.}(2005)\citenamefont {Zhang},
  \citenamefont {Tan}, \citenamefont {Stormer},\ and\ \citenamefont
  {Kim}}]{Zhang05}%
  \BibitemOpen
  \bibfield  {author} {\bibinfo {author} {\bibfnamefont {Y.}~\bibnamefont
  {Zhang}}, \bibinfo {author} {\bibfnamefont {Y.-W.}\ \bibnamefont {Tan}},
  \bibinfo {author} {\bibfnamefont {H.~L.}\ \bibnamefont {Stormer}}, \ and\
  \bibinfo {author} {\bibfnamefont {P.}~\bibnamefont {Kim}},\ }\href {\doibase
  http://www.nature.com/nature/journal/v438/n7065/suppinfo/nature04235_S1.html}
  {\bibfield  {journal} {\bibinfo  {journal} {Nature}\ }\textbf {\bibinfo
  {volume} {438}},\ \bibinfo {pages} {201} (\bibinfo {year}
  {2005})}\BibitemShut {NoStop}%
\bibitem [{\citenamefont {Lee}\ and\ \citenamefont
  {Ramakrishnan}(1985)}]{Lee85}%
  \BibitemOpen
  \bibfield  {author} {\bibinfo {author} {\bibfnamefont {P.~A.}\ \bibnamefont
  {Lee}}\ and\ \bibinfo {author} {\bibfnamefont {T.~V.}\ \bibnamefont
  {Ramakrishnan}},\ }\href {\doibase 10.1103/RevModPhys.57.287} {\bibfield
  {journal} {\bibinfo  {journal} {Rev. Mod. Phys.}\ }\textbf {\bibinfo {volume}
  {57}},\ \bibinfo {pages} {287} (\bibinfo {year} {1985})}\BibitemShut
  {NoStop}%
\bibitem [{\citenamefont {Goerbig}\ \emph {et~al.}(2008)\citenamefont
  {Goerbig}, \citenamefont {Fuchs}, \citenamefont {Montambaux},\ and\
  \citenamefont {Pi\'echon}}]{Goerbig08}%
  \BibitemOpen
  \bibfield  {author} {\bibinfo {author} {\bibfnamefont {M.~O.}\ \bibnamefont
  {Goerbig}}, \bibinfo {author} {\bibfnamefont {J.-N.}\ \bibnamefont {Fuchs}},
  \bibinfo {author} {\bibfnamefont {G.}~\bibnamefont {Montambaux}}, \ and\
  \bibinfo {author} {\bibfnamefont {F.}~\bibnamefont {Pi\'echon}},\ }\href
  {\doibase 10.1103/PhysRevB.78.045415} {\bibfield  {journal} {\bibinfo
  {journal} {Phys. Rev. B}\ }\textbf {\bibinfo {volume} {78}},\ \bibinfo
  {pages} {045415} (\bibinfo {year} {2008})}\BibitemShut {NoStop}%
\bibitem [{\citenamefont {Plihal}\ and\ \citenamefont
  {Maradudin}(1991)}]{Plihal91}%
  \BibitemOpen
  \bibfield  {author} {\bibinfo {author} {\bibfnamefont {M.}~\bibnamefont
  {Plihal}}\ and\ \bibinfo {author} {\bibfnamefont {A.~A.}\ \bibnamefont
  {Maradudin}},\ }\href {\doibase 10.1103/PhysRevB.44.8565} {\bibfield
  {journal} {\bibinfo  {journal} {Phys. Rev. B}\ }\textbf {\bibinfo {volume}
  {44}},\ \bibinfo {pages} {8565} (\bibinfo {year} {1991})}\BibitemShut
  {NoStop}%
\bibitem [{\citenamefont {Peleg}\ \emph {et~al.}(2007)\citenamefont {Peleg},
  \citenamefont {Bartal}, \citenamefont {Freedman}, \citenamefont {Manela},
  \citenamefont {Segev},\ and\ \citenamefont {Christodoulides}}]{Peleg07}%
  \BibitemOpen
  \bibfield  {author} {\bibinfo {author} {\bibfnamefont {O.}~\bibnamefont
  {Peleg}}, \bibinfo {author} {\bibfnamefont {G.}~\bibnamefont {Bartal}},
  \bibinfo {author} {\bibfnamefont {B.}~\bibnamefont {Freedman}}, \bibinfo
  {author} {\bibfnamefont {O.}~\bibnamefont {Manela}}, \bibinfo {author}
  {\bibfnamefont {M.}~\bibnamefont {Segev}}, \ and\ \bibinfo {author}
  {\bibfnamefont {D.~N.}\ \bibnamefont {Christodoulides}},\ }\href {\doibase
  10.1103/PhysRevLett.98.103901} {\bibfield  {journal} {\bibinfo  {journal}
  {Phys. Rev. Lett.}\ }\textbf {\bibinfo {volume} {98}},\ \bibinfo {pages}
  {103901} (\bibinfo {year} {2007})}\BibitemShut {NoStop}%
\bibitem [{\citenamefont {Wang}\ \emph {et~al.}(2009)\citenamefont {Wang},
  \citenamefont {Wang}, \citenamefont {Zhang},\ and\ \citenamefont
  {Zhu}}]{Wang09}%
  \BibitemOpen
  \bibfield  {author} {\bibinfo {author} {\bibfnamefont {L.-G.}\ \bibnamefont
  {Wang}}, \bibinfo {author} {\bibfnamefont {Z.-G.}\ \bibnamefont {Wang}},
  \bibinfo {author} {\bibfnamefont {J.-X.}\ \bibnamefont {Zhang}}, \ and\
  \bibinfo {author} {\bibfnamefont {S.-Y.}\ \bibnamefont {Zhu}},\ }\href
  {\doibase 10.1364/OL.34.001510} {\bibfield  {journal} {\bibinfo  {journal}
  {Opt. Lett.}\ }\textbf {\bibinfo {volume} {34}},\ \bibinfo {pages} {1510}
  (\bibinfo {year} {2009})}\BibitemShut {NoStop}%
\bibitem [{\citenamefont {Huang}\ \emph {et~al.}(2011)\citenamefont {Huang},
  \citenamefont {Lai}, \citenamefont {Hang}, \citenamefont {Zheng},\ and\
  \citenamefont {Chan}}]{Huang11}%
  \BibitemOpen
  \bibfield  {author} {\bibinfo {author} {\bibfnamefont {X.}~\bibnamefont
  {Huang}}, \bibinfo {author} {\bibfnamefont {Y.}~\bibnamefont {Lai}}, \bibinfo
  {author} {\bibfnamefont {Z.~H.}\ \bibnamefont {Hang}}, \bibinfo {author}
  {\bibfnamefont {H.}~\bibnamefont {Zheng}}, \ and\ \bibinfo {author}
  {\bibfnamefont {C.~T.}\ \bibnamefont {Chan}},\ }\href@noop {} {\bibfield
  {journal} {\bibinfo  {journal} {Nat Mater}\ }\textbf {\bibinfo {volume}
  {10}},\ \bibinfo {pages} {582} (\bibinfo {year} {2011})}\BibitemShut
  {NoStop}%
\bibitem [{\citenamefont {Sun}\ \emph {et~al.}(2013{\natexlab{b}})\citenamefont
  {Sun}, \citenamefont {Gao},\ and\ \citenamefont {Yang}}]{LSun13}%
  \BibitemOpen
  \bibfield  {author} {\bibinfo {author} {\bibfnamefont {L.}~\bibnamefont
  {Sun}}, \bibinfo {author} {\bibfnamefont {J.}~\bibnamefont {Gao}}, \ and\
  \bibinfo {author} {\bibfnamefont {X.}~\bibnamefont {Yang}},\ }\href {\doibase
  10.1364/OE.21.021542} {\bibfield  {journal} {\bibinfo  {journal} {Opt.
  Express}\ }\textbf {\bibinfo {volume} {21}},\ \bibinfo {pages} {21542}
  (\bibinfo {year} {2013}{\natexlab{b}})}\BibitemShut {NoStop}%
\bibitem [{\citenamefont {Born}\ \emph {et~al.}(1999)\citenamefont {Born},
  \citenamefont {Wolf},\ and\ \citenamefont {Bhatia}}]{Born99}%
  \BibitemOpen
  \bibfield  {author} {\bibinfo {author} {\bibfnamefont {M.}~\bibnamefont
  {Born}}, \bibinfo {author} {\bibfnamefont {E.}~\bibnamefont {Wolf}}, \ and\
  \bibinfo {author} {\bibfnamefont {A.}~\bibnamefont {Bhatia}},\ }\href
  {http://books.google.ie/books?id=aoX0gYLuENoC} {\emph {\bibinfo {title}
  {Principles of Optics: Electromagnetic Theory of Propagation, Interference
  and Diffraction of Light}}}\ (\bibinfo  {publisher} {Cambridge University
  Press},\ \bibinfo {year} {1999})\BibitemShut {NoStop}%
\bibitem [{\citenamefont {Niklasson}\ \emph {et~al.}(1981)\citenamefont
  {Niklasson}, \citenamefont {Granqvist},\ and\ \citenamefont
  {Hunderi}}]{Niklasson81}%
  \BibitemOpen
  \bibfield  {author} {\bibinfo {author} {\bibfnamefont {G.~A.}\ \bibnamefont
  {Niklasson}}, \bibinfo {author} {\bibfnamefont {C.~G.}\ \bibnamefont
  {Granqvist}}, \ and\ \bibinfo {author} {\bibfnamefont {O.}~\bibnamefont
  {Hunderi}},\ }\href {\doibase 10.1364/AO.20.000026} {\bibfield  {journal}
  {\bibinfo  {journal} {Appl. Opt.}\ }\textbf {\bibinfo {volume} {20}},\
  \bibinfo {pages} {26} (\bibinfo {year} {1981})}\BibitemShut {NoStop}%
\bibitem [{\citenamefont {Landau}\ \emph {et~al.}(1984)\citenamefont {Landau},
  \citenamefont {Lifshitz},\ and\ \citenamefont {Pitaevskii}}]{Landau84}%
  \BibitemOpen
  \bibfield  {author} {\bibinfo {author} {\bibfnamefont {L.}~\bibnamefont
  {Landau}}, \bibinfo {author} {\bibfnamefont {E.}~\bibnamefont {Lifshitz}}, \
  and\ \bibinfo {author} {\bibfnamefont {L.}~\bibnamefont {Pitaevskii}},\
  }\href@noop {} {\emph {\bibinfo {title} {Electrodynamics of continuous
  media}}}\ (\bibinfo  {publisher} {Pergamon},\ \bibinfo {year}
  {1984})\BibitemShut {NoStop}%
\bibitem [{\citenamefont {Hecht}(2002)}]{Hecht02}%
  \BibitemOpen
  \bibfield  {author} {\bibinfo {author} {\bibfnamefont {E.}~\bibnamefont
  {Hecht}},\ }\href {http://books.google.ie/books?id=7aG6QgAACAAJ} {\emph
  {\bibinfo {title} {Optics}}}\ (\bibinfo  {publisher} {Addison-Wesley Longman,
  Incorporated},\ \bibinfo {year} {2002})\BibitemShut {NoStop}%
\bibitem [{\citenamefont {{Gao}}\ \emph {et~al.}(2014)\citenamefont {{Gao}},
  \citenamefont {{Lawrence}}, \citenamefont {{Yang}}, \citenamefont {{Liu}},
  \citenamefont {{Fang}}, \citenamefont {{Li}},\ and\ \citenamefont
  {{Zhang}}}]{Gao14}%
  \BibitemOpen
  \bibfield  {author} {\bibinfo {author} {\bibfnamefont {W.}~\bibnamefont
  {{Gao}}}, \bibinfo {author} {\bibfnamefont {M.}~\bibnamefont {{Lawrence}}},
  \bibinfo {author} {\bibfnamefont {B.}~\bibnamefont {{Yang}}}, \bibinfo
  {author} {\bibfnamefont {F.}~\bibnamefont {{Liu}}}, \bibinfo {author}
  {\bibfnamefont {F.}~\bibnamefont {{Fang}}}, \bibinfo {author} {\bibfnamefont
  {J.}~\bibnamefont {{Li}}}, \ and\ \bibinfo {author} {\bibfnamefont
  {S.}~\bibnamefont {{Zhang}}},\ }\href@noop {} {\bibfield  {journal} {\bibinfo
   {journal} {ArXiv e-prints}\ } (\bibinfo {year} {2014})},\ \Eprint
  {http://arxiv.org/abs/1401.5448} {arXiv:1401.5448 [physics.optics]}
  \BibitemShut {NoStop}%
\bibitem [{\citenamefont {de~Groot}\ \emph {et~al.}(1983)\citenamefont
  {de~Groot}, \citenamefont {Mueller}, \citenamefont {vanEngen},\ and\
  \citenamefont {Buschow}}]{Groot83}%
  \BibitemOpen
  \bibfield  {author} {\bibinfo {author} {\bibfnamefont {R.~A.}\ \bibnamefont
  {de~Groot}}, \bibinfo {author} {\bibfnamefont {F.~M.}\ \bibnamefont
  {Mueller}}, \bibinfo {author} {\bibfnamefont {P.~G.~v.}\ \bibnamefont
  {vanEngen}}, \ and\ \bibinfo {author} {\bibfnamefont {K.~H.~J.}\ \bibnamefont
  {Buschow}},\ }\href {\doibase 10.1103/PhysRevLett.50.2024} {\bibfield
  {journal} {\bibinfo  {journal} {Phys. Rev. Lett.}\ }\textbf {\bibinfo
  {volume} {50}},\ \bibinfo {pages} {2024} (\bibinfo {year}
  {1983})}\BibitemShut {NoStop}%
\bibitem [{\citenamefont {Khanikaev}\ \emph {et~al.}(2013)\citenamefont
  {Khanikaev}, \citenamefont {Hossein~Mousavi}, \citenamefont {Tse},
  \citenamefont {Kargarian}, \citenamefont {MacDonald},\ and\ \citenamefont
  {Shvets}}]{Khanikaev13}%
  \BibitemOpen
  \bibfield  {author} {\bibinfo {author} {\bibfnamefont {A.~B.}\ \bibnamefont
  {Khanikaev}}, \bibinfo {author} {\bibfnamefont {S.}~\bibnamefont
  {Hossein~Mousavi}}, \bibinfo {author} {\bibfnamefont {W.-K.}\ \bibnamefont
  {Tse}}, \bibinfo {author} {\bibfnamefont {M.}~\bibnamefont {Kargarian}},
  \bibinfo {author} {\bibfnamefont {A.~H.}\ \bibnamefont {MacDonald}}, \ and\
  \bibinfo {author} {\bibfnamefont {G.}~\bibnamefont {Shvets}},\ }\href@noop {}
  {\bibfield  {journal} {\bibinfo  {journal} {Nat Mater}\ }\textbf {\bibinfo
  {volume} {12}},\ \bibinfo {pages} {233} (\bibinfo {year} {2013})}\BibitemShut
  {NoStop}%
\bibitem [{\citenamefont {Dragoman}\ and\ \citenamefont
  {Dragoman}(2004)}]{Dragoman04}%
  \BibitemOpen
  \bibfield  {author} {\bibinfo {author} {\bibfnamefont {D.}~\bibnamefont
  {Dragoman}}\ and\ \bibinfo {author} {\bibfnamefont {M.}~\bibnamefont
  {Dragoman}},\ }\href@noop {} {\emph {\bibinfo {title} {Quantum-Classical
  Analogies}}},\ The Frontiers Collection\ (\bibinfo  {publisher} {Springer},\
  \bibinfo {year} {2004})\BibitemShut {NoStop}%
\bibitem [{\citenamefont {Haldane}(1988)}]{Haldane88}%
  \BibitemOpen
  \bibfield  {author} {\bibinfo {author} {\bibfnamefont {F.~D.~M.}\
  \bibnamefont {Haldane}},\ }\href {\doibase 10.1103/PhysRevLett.61.2015}
  {\bibfield  {journal} {\bibinfo  {journal} {Phys. Rev. Lett.}\ }\textbf
  {\bibinfo {volume} {61}},\ \bibinfo {pages} {2015} (\bibinfo {year}
  {1988})}\BibitemShut {NoStop}%
\bibitem [{\citenamefont {Kane}\ and\ \citenamefont {Mele}(2005)}]{Kane05}%
  \BibitemOpen
  \bibfield  {author} {\bibinfo {author} {\bibfnamefont {C.~L.}\ \bibnamefont
  {Kane}}\ and\ \bibinfo {author} {\bibfnamefont {E.~J.}\ \bibnamefont
  {Mele}},\ }\href {\doibase 10.1103/PhysRevLett.95.146802} {\bibfield
  {journal} {\bibinfo  {journal} {Phys. Rev. Lett.}\ }\textbf {\bibinfo
  {volume} {95}},\ \bibinfo {pages} {146802} (\bibinfo {year}
  {2005})}\BibitemShut {NoStop}%
\bibitem [{\citenamefont {Hasan}\ and\ \citenamefont {Kane}(2010)}]{Hasan10}%
  \BibitemOpen
  \bibfield  {author} {\bibinfo {author} {\bibfnamefont {M.~Z.}\ \bibnamefont
  {Hasan}}\ and\ \bibinfo {author} {\bibfnamefont {C.~L.}\ \bibnamefont
  {Kane}},\ }\href {\doibase 10.1103/RevModPhys.82.3045} {\bibfield  {journal}
  {\bibinfo  {journal} {Rev. Mod. Phys.}\ }\textbf {\bibinfo {volume} {82}},\
  \bibinfo {pages} {3045} (\bibinfo {year} {2010})}\BibitemShut {NoStop}%
\bibitem [{\citenamefont {Raghu}\ and\ \citenamefont
  {Haldane}(2008)}]{Raghu08}%
  \BibitemOpen
  \bibfield  {author} {\bibinfo {author} {\bibfnamefont {S.}~\bibnamefont
  {Raghu}}\ and\ \bibinfo {author} {\bibfnamefont {F.~D.~M.}\ \bibnamefont
  {Haldane}},\ }\href {\doibase 10.1103/PhysRevA.78.033834} {\bibfield
  {journal} {\bibinfo  {journal} {Phys. Rev. A}\ }\textbf {\bibinfo {volume}
  {78}},\ \bibinfo {pages} {033834} (\bibinfo {year} {2008})}\BibitemShut
  {NoStop}%
\bibitem [{\citenamefont {Ochiai}\ and\ \citenamefont
  {Onoda}(2009)}]{Ochiai09}%
  \BibitemOpen
  \bibfield  {author} {\bibinfo {author} {\bibfnamefont {T.}~\bibnamefont
  {Ochiai}}\ and\ \bibinfo {author} {\bibfnamefont {M.}~\bibnamefont {Onoda}},\
  }\href {\doibase 10.1103/PhysRevB.80.155103} {\bibfield  {journal} {\bibinfo
  {journal} {Phys. Rev. B}\ }\textbf {\bibinfo {volume} {80}},\ \bibinfo
  {pages} {155103} (\bibinfo {year} {2009})}\BibitemShut {NoStop}%
\bibitem [{\citenamefont {Rechtsman}\ \emph
  {et~al.}(2013{\natexlab{a}})\citenamefont {Rechtsman}, \citenamefont
  {Plotnik}, \citenamefont {Zeuner}, \citenamefont {Song}, \citenamefont
  {Chen}, \citenamefont {Szameit},\ and\ \citenamefont {Segev}}]{Rechtsman13b}%
  \BibitemOpen
  \bibfield  {author} {\bibinfo {author} {\bibfnamefont {M.~C.}\ \bibnamefont
  {Rechtsman}}, \bibinfo {author} {\bibfnamefont {Y.}~\bibnamefont {Plotnik}},
  \bibinfo {author} {\bibfnamefont {J.~M.}\ \bibnamefont {Zeuner}}, \bibinfo
  {author} {\bibfnamefont {D.}~\bibnamefont {Song}}, \bibinfo {author}
  {\bibfnamefont {Z.}~\bibnamefont {Chen}}, \bibinfo {author} {\bibfnamefont
  {A.}~\bibnamefont {Szameit}}, \ and\ \bibinfo {author} {\bibfnamefont
  {M.}~\bibnamefont {Segev}},\ }\href {\doibase 10.1103/PhysRevLett.111.103901}
  {\bibfield  {journal} {\bibinfo  {journal} {Phys. Rev. Lett.}\ }\textbf
  {\bibinfo {volume} {111}},\ \bibinfo {pages} {103901} (\bibinfo {year}
  {2013}{\natexlab{a}})}\BibitemShut {NoStop}%
\bibitem [{\citenamefont {Rechtsman}\ \emph
  {et~al.}(2013{\natexlab{b}})\citenamefont {Rechtsman}, \citenamefont
  {Zeuner}, \citenamefont {Plotnik}, \citenamefont {Lumer}, \citenamefont
  {Podolsky}, \citenamefont {Dreisow}, \citenamefont {Nolte}, \citenamefont
  {Segev},\ and\ \citenamefont {Szameit}}]{Rechtsman13}%
  \BibitemOpen
  \bibfield  {author} {\bibinfo {author} {\bibfnamefont {M.~C.}\ \bibnamefont
  {Rechtsman}}, \bibinfo {author} {\bibfnamefont {J.~M.}\ \bibnamefont
  {Zeuner}}, \bibinfo {author} {\bibfnamefont {Y.}~\bibnamefont {Plotnik}},
  \bibinfo {author} {\bibfnamefont {Y.}~\bibnamefont {Lumer}}, \bibinfo
  {author} {\bibfnamefont {D.}~\bibnamefont {Podolsky}}, \bibinfo {author}
  {\bibfnamefont {F.}~\bibnamefont {Dreisow}}, \bibinfo {author} {\bibfnamefont
  {S.}~\bibnamefont {Nolte}}, \bibinfo {author} {\bibfnamefont
  {M.}~\bibnamefont {Segev}}, \ and\ \bibinfo {author} {\bibfnamefont
  {A.}~\bibnamefont {Szameit}},\ }\href {\doibase 10.1038/nature12066
  http://www.nature.com/nature/journal/v496/n7444/abs/nature12066.html#supplementary-information}
  {\bibfield  {journal} {\bibinfo  {journal} {Nature}\ }\textbf {\bibinfo
  {volume} {496}},\ \bibinfo {pages} {196} (\bibinfo {year}
  {2013}{\natexlab{b}})}\BibitemShut {NoStop}%
\end{thebibliography}
\end{document}